\documentclass[11pt]{article}

\usepackage[preprint]{acl}

\usepackage{times}
\usepackage{latexsym}

\usepackage[T1]{fontenc}

\usepackage[utf8]{inputenc}

\usepackage{microtype}

\usepackage{inconsolata}

\usepackage{graphicx}

\usepackage{CJKutf8} 

\usepackage[most]{tcolorbox}
\usepackage{helvet} 
\usepackage{etoolbox}
\tcbuselibrary{skins}  

\definecolor{qtBg}{HTML}{E6F2EF}\definecolor{qtFg}{HTML}{2E7D71}
\definecolor{qlBg}{HTML}{F3E8F1}\definecolor{qlFg}{HTML}{8E4585}
\definecolor{tcBg}{HTML}{E7EEF6}\definecolor{tcFg}{HTML}{3A6EA5} 
\definecolor{crBg}{HTML}{F8E9E5}\definecolor{crFg}{HTML}{C05746}
\definecolor{boxBg}{HTML}{F6F6F7}
\definecolor{boxLine}{HTML}{E3E3E6}

\newcommand{\sspill}[3]{\tikz[baseline=(x.base)]{\node[inner xsep=4pt,
  inner ysep=1.7pt, rounded corners=2.5pt, fill=#1, text=#2,
  font=\fontsize{6.6}{7}\selectfont\bfseries\sffamily] (x) {#3};}}
\definecolor{usBg}{HTML}{E7E9F7}\definecolor{usFg}{HTML}{3A4FA8} 
\definecolor{fsBg}{HTML}{E3EFF6}\definecolor{fsFg}{HTML}{2A7EA8}
\definecolor{mcBg}{HTML}{E6F2EF}\definecolor{mcFg}{HTML}{2E7D71}
\definecolor{bmBg}{HTML}{EAF3E4}\definecolor{bmFg}{HTML}{4F8A3D} 
\definecolor{sbBg}{HTML}{F6EEDC}\definecolor{sbFg}{HTML}{A6731F}
\definecolor{pdBg}{HTML}{F3E8F1}\definecolor{pdFg}{HTML}{8E4585} 
\definecolor{fbBg}{HTML}{F8E9E5}\definecolor{fbFg}{HTML}{C05746} 
\newcommand{\iUser}{\sspill{usBg}{usFg}{USER STUDIES}}
\newcommand{\iField}{\sspill{fsBg}{fsFg}{FIELD STUDIES}}
\newcommand{\iMetric}{\sspill{mcBg}{mcFg}{METRIC CREATION}}
\newcommand{\iBench}{\sspill{bmBg}{bmFg}{BENCHMARKING}}
\newcommand{\iSystem}{\sspill{sbBg}{sbFg}{SYSTEM BUILDING}}
\newcommand{\iPartic}{\sspill{pdBg}{pdFg}{PARTICIPATORY DESIGN}}
\newcommand{\iFieldbuild}{\sspill{fbBg}{fbFg}{FIELD-BUILDING}}
\newcommand{\tUser}{\iUser\hspace{2.5pt}}
\newcommand{\tField}{\iField\hspace{2.5pt}}
\newcommand{\tMetric}{\iMetric\hspace{2.5pt}}
\newcommand{\tBench}{\iBench\hspace{2.5pt}}
\newcommand{\tSystem}{\iSystem\hspace{2.5pt}}
\newcommand{\tPartic}{\iPartic\hspace{2.5pt}}
\newcommand{\tFieldbuild}{\iFieldbuild\hspace{2.5pt}}

\newtcolorbox{solutionsketch}[2][]{%
  enhanced, breakable, colback=boxBg, colframe=boxLine, boxrule=0.8pt, arc=4pt,
  left=9pt, right=9pt, top=5pt, bottom=6pt,
  drop fuzzy shadow={black!18!white},
  before upper={%
    \setlength{\parskip}{0pt}\setlength{\parindent}{0pt}%
    {\bfseries\ifstrempty{#1}{}{Solution Sketch #1}}\par\vspace{-1.5pt}%
    \noindent#2\par\vspace{1pt}%
  },
}

\newcommand{\caveat}[1]{\par\vspace{2pt}\noindent%
  {\footnotesize\itshape\textbf{Caveat:}\space#1}\par}

\title{A Cross-Community Agenda for Speech AI}

\author{
  \textbf{Maria Teleki\textsuperscript{1,*}},
  \textbf{Kimi V. Wenzel\textsuperscript{2,*}},
  \textbf{Anna Seo Gyeong Choi\textsuperscript{3,*}},
    \textbf{Tobias Weinberg\textsuperscript{4,*}},
\\
  \textbf{Shree Harsha Bokkahalli Satish\textsuperscript{5,*}},
  \textbf{Stephanny Sanchez\textsuperscript{6,*}},
    \textbf{Belu Ticona\textsuperscript{7,*}},
\\
  \textbf{Ariadna Sanchez\textsuperscript{8,*}},
    \textbf{Yash Sonkar\textsuperscript{9,*}},
    \textbf{Aarti Mathur\textsuperscript{10,*}},
 \textbf{Christoph Minixhofer\textsuperscript{11,*}},
\\
  \textbf{Abraham Glasser\textsuperscript{12,*}},
  \textbf{Raja Kushalnagar\textsuperscript{12,*}},
      \textbf{James Caverlee\textsuperscript{1,*}},
    \textbf{Minha Lee\textsuperscript{13,*}},
\\
    \textbf{Shaomei Wu\textsuperscript{14,*}},
    \textbf{Alyssa Hillary Zisk\textsuperscript{15,*}},
  \textbf{Éva Székely\textsuperscript{5,*}},
      \textbf{Dylan Gaines\textsuperscript{16,*}},
\\
    \textbf{Angelika Seeschaaf Veres\textsuperscript{17,*}},
    \textbf{Seray Ibrahim\textsuperscript{18,*}},
  \textbf{Nicholas Cummins\textsuperscript{18,*}},
  \textbf{Allison Koenecke\textsuperscript{3,4,*}}
\\
\\
  \textsuperscript{1}Texas A\&M University,
  \textsuperscript{2}Carnegie Mellon University,
  \textsuperscript{3}Cornell University,
    \textsuperscript{4}Cornell Tech,\\
      \textsuperscript{5}KTH Royal Institute of Technology,
  \textsuperscript{6}University of Alberta,
    \textsuperscript{7}George Mason University\\
      \textsuperscript{8}University of Edinburgh,
      \textsuperscript{9}IIIT Hyderabad,
  \textsuperscript{10}Chegg Inc.,
      \textsuperscript{11}Deepgram,\\
  \textsuperscript{12}Gallaudet University,
    \textsuperscript{13}Eindhoven University of Technology,\\
      \textsuperscript{14}AImpower.org,
  \textsuperscript{15}AssistiveWare,
  \textsuperscript{16}Kennesaw State University,\\
    \textsuperscript{17}OCAD University,
  \textsuperscript{18}King's College London
\\
\\
  \small{
    \textbf{* All authors contributed equally.}
    \textbf{Correspondence:} \texttt{mariateleki@tamu.edu, kwenzel@andrew.cmu.edu}
  }
}

\begin{document}
\maketitle
\begin{abstract}
Speech AI, any AI system that recognizes, transforms, or generates speech, is built and evaluated across two communities with only a small overlap: technical natural language processing (NLP) venues (e.g., *ACL, ICASSP, Interspeech), and sociotechnical HCI venues (e.g., ASSETS, CHI, FAccT). In this position paper, we work toward a cross-community synthesis, organizing our critique around three problems: speech AI operates with an incomplete model of communication; it operates with an incomplete model of identity; and its metrics measure the wrong constructs. We draw on AAC as a setting where these failures are most visible and their stakes highest, alongside other underserved speakers — people who stutter, multilingual speakers, and non-binary and transgender users. For each problem we offer solution sketches oriented toward designing for human variability, nearly all of which require quantitative and qualitative methods in combination. We close on the venue structures that hold these methods apart, and on what program committees and individual authors can do to bring them together.
\end{abstract}

\section*{Introduction}

Speech AI\footnote{In this work, we use \textit{speech AI} to refer to the range of systems that recognize, transform, or generate speech, at any level from the individual component to the deployed application. This includes single components operating on their own --- ASR transcribing speech to text, TTS synthesizing text into speech, and voice conversion transforming one voice into another. It also includes \textit{chained} pipelines, where an ASR module transcribes speech to text, an LLM operates on that text, and a TTS module synthesizes the response back into speech with each stage handled by a separate model. Further examples of speech AI are newer \textit{full-duplex} spoken dialogue models, which process and generate speech directly and continuously in both directions, without discrete turn-taking between separate modules. Applications built on these components include AAC devices, voice assistants, captioning services, and clinical scribes. See reviews on clinical speech AI \cite{ng2026end} and SpeechLLMs \cite{arora2025landscape, cui2025recent, yang2025large}.} is built and evaluated in two places at once. Technical venues like *ACL, ICASSP, and Interspeech produce the models, the corpora, and the metrics by which progress is declared. Sociotechnical venues like ASSETS, CHI, and FAccT produce the accounts of how people use these systems and corresponding desiderata. The two communities examine the same artifacts, and they overlap only thinly in citation practice, method, and reviewer pool. This position paper argues that the most consequential open problems in speech AI sit in the gap between them, and that closing those problems requires work that neither community can carry out alone.

Speech makes this divide costly in a way that other modalities do not, because speech is uniquely indexical. Unlike text, spoken language \textit{unavoidably} encodes social identity --- accent, intonation, speech style, and prosody signal group affiliation, personality, health status, and stance in ways that written language does not \cite{eckert2019limits, szekely2025will, foulkes2006social, eckert2008variation, teleki25_icwsm, weinberg_one_2025}. A system that recognizes or generates speech is therefore always doing two things
simultaneously: performing a measurable technical operation, and making a social judgment about a speaker. 

This divide is apparent in how both communities negotiate diversity and fairness. Both communities invoke `diversity' consistently, yet the term indexes fundamentally different commitments within each research community. In technical venues, diversity is operationalized as demographic breadth in training corpora, (e.g. multiple accents, languages, or speaker age groups), and its adequacy is judged by whether a system generalizes across a predefined taxonomy
of speaker categories, evaluated through aggregate metrics disaggregated by group \cite{dheram2022toward, liu2022model, papakyriakopoulos2023augmented}. Diversity, in this framing, is a property of \textit{data}. Where atypical speech enters at all, it does so predominantly within a pathological frame, as something to be detected, corrected, or accommodated through better data augmentation \cite{tobin2025towards, tomanek2023analysis}. The social meaning of a voice is never fully captured by the demographic category it is assigned to, and designing for diversity without accounting for this indexical richness reproduces the very exclusions it claims to remedy.

HCI communities treat diversity instead as a property of \textit{experience}: who is heard, on whose terms, and with what consequences for identity, self-expression, and participation \cite{erete2018intersectional, ibrahim_design_2018, wu2025speech, michel2025bias, weinberg2026robot}. Being represented in a corpus and being served by a system are not equivalent, and conflating them has attracted sustained critique under the label of `diversity-washing': the risk that superficial inclusion efforts obscure structural inequities or bypass community consent \cite{cunningham2025advancing}. 

The NLP community has begun to grapple with this. Recent work argues that diversity in language technology is an ambiguous and underspecified concept admitting potentially conflicting operationalizations, and calls for the field to explicitly negotiate what diversity \textit{means} before debating how to measure it \cite{fazelpour2022diversity, nguyen2025we, bowman2021will}. Earlier advances in the text and vision spaces included critical surveys interrogating the vocabulary of fairness itself \cite{blodgett2021stereotyping, blodgett2020language}, and audits that made subgroup disparity legible and established disaggregated and extrinsic evaluation as an expectation \cite{goldfarb2021intrinsic, buolamwini2018gender}. While analogous inroads have been made in the speech modality, they have often temporally lagged other domains (e.g., the \citet{koenecke2020racial} ASR audit was directly inspired by the \citet{buolamwini2018gender} facial recognition audit). Speech AI has yet to mount a comparable reckoning in defining diversity. Some work has begun to bring philosophical scrutiny to bear on what fairness means specifically for ASR \cite{choi2025fairness}, rather than importing the concept unexamined from adjacent modalities. Cross-community translation between technical and sociotechnical speech venues is not simply a matter of citation practices; it requires first agreeing on a shared agenda.

In this position paper, we set an agenda for \textbf{a cross-community synthesis}, across technical and sociotechnical venues. We present \textbf{a critique of both speech AI communities organized around three problems:} 

\begin{description}
    \item[\textbf{Problem 1:}] Speech AI operates with an incomplete model of communication.
    \item[\textbf{Problem 2:}] Speech AI operates with an incomplete model of identity.
    \item[\textbf{Problem 3:}] Speech AI metrics measure the wrong constructs.
\end{description}

For each of these three failures, we present \textbf{a set of solution sketches}\footnote{Similarly to \citet{bowman2021will}.} in Speech AI that center human variability, offering a path forward for both scientific communities. We ground each problem and solution sketch in a case study of augmentative and alternative communication (AAC),\footnote{Definitions of AAC vary, but generally cover the use of communication methods other than speech for disability-related reasons. AAC supports include communication boards, books, devices, and more; speech AI is relevant to AAC systems with speech output. For AAC users, their AAC is a large part of their identity and presentation \cite{Tuttleturtle2020gender}, making speech AI an important issue for many AAC users.} as this is a speech AI use case where issues are particularly salient and consequential to users. To help researchers see where they can contribute, we tag each solution sketch by the kind of work it invites --- from empirical study (\iUser, \iField) and measurement (\iMetric, \iBench) to system building (\iSystem) and community practice (\iPartic, \iFieldbuild); detailed descriptions of each tag are provided in Appendix~\ref{appendix-contributor-tags}. Most sketches carry more than one tag, because progress on these problems requires a multi-pronged approach. 

\section*{Problem 1: Speech AI operates with an incomplete model of communication.}
We begin with the model of communication that conversational speech AI implicitly assumes, since the commitments made there propagate into how these systems represent identity and how the field measures its own progress.
A core idea in contemporary studies of social interaction is that speech is just one ingredient of communication \cite{levinson_interaction_2025}, deeply nestled within a wider ecology of multimodal streams, cooperative acts, and co-constructed practices of two or more parties \cite{clark_using_1996, goodwin_conversation_1990}. Wider streams that impact speech and communication can include suprasegmental features (e.g., stress, intonation, rhythm), non-spoken modes that nonetheless convey linguistic information in signed languages (e.g., facial expression, gaze, gesture) \cite{hill2020deaf}, and importantly, collaborative patterns between interactants to establish mutual understanding \cite{goodwin_action_2000}. When deciding how we might investigate or design potential speech AI supports, it is important to acknowledge these wider processes. 

These dynamics are constitutive of communication; active-listening signals, turn-taking cues, humor, and embodied feedback are part of how speakers assert presence, negotiate meaning, and remain recognizable as themselves across an interaction~\cite{weinberg2025why, weinberg_one_2025, jurafsky1997automatic}. Current speech AI systems, even full-duplex models, largely treat these as noise or omit them entirely, optimizing instead for the verbal channel alone. This is not merely a coverage gap; it reflects a fundamentally impoverished model of what communication is.

Considering speech AI for AAC, numerous prior studies have shown that speech generation via AAC devices can produce understandability issues between interactants \cite{bloch_understandability_2004, higginbotham2002aac, ibrahim_design_2018}. These wider streams and conversational practices impact the perception of AAC-generated speech during social interactions and how far AAC users can agentively express themselves during conversations \cite{ valencia_conversational_2020, weinberg_one_2025}. 
For AAC users in particular, the inability to produce timely non-verbal cues --- whether a backchannel, an expressive reaction, or a well-timed humorous remark --- is not only a conversational inconvenience but a constraint on identity expression itself~\cite{weinberg2025why, weinberg_one_2025}. However, a major focus of speech AI for AAC has been on enhancing speech efficiency with less focus on interactional issues such as reciprocity, conversational management, or identity expression \cite{ibrahim_before_2026}. We highlight issues of identity expression directly in Problem 2.

\begin{solutionsketch}[1]{\tSystem\tUser\tPartic}
Future speech AI models should include other dimensions of communication such as, but not limited to, emotional expression. Emotion models fit to third-party labels transfer poorly to self-reported emotion \cite{el2026you}, and closing that gap requires models personalized to the individual \cite{tavernor2026personal}; we take this as the design problem for expressive AAC output. A personalized expressive model, trained on data the user contributes and governs, would give an AAC user a vocabulary of prosodic states they recognize as their own and can deploy deliberately in conversation. Facial expression and physiological data from a wearable can serve as inputs for users who opt into them, on the terms those users set. Evaluation follows the same logic; the user reports what they intended to convey, communication partners report what they received, and the gap between these is examined under success criteria the user defines.
\caveat{Sensing of this kind carries documented bias \cite{holliday2025gender}. Personalization without user control over the model and its data collapses into surveillance, and an expressive system that optimizes a physiological proxy
reproduces the failure mode described in Problem 3.}
\end{solutionsketch}

\section*{Problem 2: Speech AI operates with an incomplete model of identity.}

We attribute speech AI's incomplete identity model to reductive user models (Problem 2A) and inaccurate representation for speech synthesis users (Problem 2B). Both of these issues feed back into Problem 1, insofar as speech AI's model of communication lacks appropriate recognition and operationalization of user identity. 

\subsection*{Problem 2A: Speech AI user models are reductive.}

Technologies have co-produced language ideologies in which users are expected to fit certain taxonomies, and variation is treated as pathology instead of data. Models that treat `atypical' users as pathological, ignorable, or treatable with sufficient data augmentation simplify away the variation that is the data, e.g., the true norm of multilingualism \cite{schneider2022multilingualism} in favor of national monolingual ideologies. Variations in vocal quality associated with identity markers \cite{wenzel-2023-microaggressors, ahmed2022app} or disability \cite{preece2024making,li2025collective} are averaged out without anyone asking the users being averaged what they think about it \cite{radford2022whisper,li2024reenvisioning}. 

Creating accurate user models is inherently challenged by human identities being neither discrete nor constant. Current taxonomies often rely on reductive frameworks that fail to capture the fluid spectrum of identities individuals embody or the intersectionality of their experiences, such as the compounding acoustic variations of gender, age, and regional dialect. An individual's voice, vocabulary, and speaking style evolve gradually with age \cite{rojas2020does}, dialect drift~\cite{harrington_acoustic_2006}, gender fluidity \cite{netzorg2024speech}, or progressive shifts in communication ability such as with dysarthria~\cite{rosen_longitudinal_2012}. They can also adapt immediately to context such as through code-switching~\cite{poplack_codeswitching_1980} or audience adaptation. Speech AI devices themselves can also impact intra-speaker variation, for example, a device enabling faster input may lead a speaker to be more verbose, while a slower one may encourage brevity.

Resisting any of this through personalization introduces its own tension. Training a model on one's own communication data can better reflect individual tone, vocabulary, and cultural identity, but raises unresolved questions of authorship, data governance, and contextual appropriateness \cite{rincon2021speaking, li2025govern, weinberg2026robot}. Technical obstacles compound the normative ones: post-training methods need data that represents individuals across their multiple identity dimensions, personalized evaluation frameworks barely exist, and many popular LLMs do not publish the model weights that post-training would need to modify.

\begin{solutionsketch}[2A]{\tPartic\tSystem}
Popular AI models are designed and owned by few companies with their own interests and agendas, for whom authentic individual user identities might not represent an interest. Grassroots communities and cross-institutional collaborations that center systematically neglected identities could foster spaces to develop new alternative ways of technology design and model development. Examples include the Masakhane community that develops AI for and by Africans \cite{nekoto2020participatory}, the Te Hiku Media Project that develops ASR for and by M\=aori people \cite{leoni-etal-2024-solving}, among others. The Mozilla Data Collective recently released a \textit{community compensation} feature, allowing communities to get paid for their data \cite{MDC2026linkedin}. Community ownership addresses who builds the model; Sketch 3C takes up the complementary question of how such a model would be evaluated across the fluid and intersectional identities described above.
\end{solutionsketch}

\subsection*{Problem 2B: Speech synthesis users need a voice that accurately represents them.}

Where Problem 2A concerned the reductive user models which form the basis of speech AI systems, Problem 2B points towards issues of system output and user experience. When examining speech synthesis output, we encounter issues of both avowed identity (how an individual identifies themself, reflexively) and ascribed identity (how others attribute identities to an individual) \cite{bucholtz_identity_2005, martin2010intercultural}.

Speech AI systems are trained on vast quantities of data drawn from many contexts, cultures, and individual authors, and the large foundational models at their core generate text and synthesize speech according to the patterns most common across that data. The result is an implicit \textit{averaging effect} wherein individual voices and identity expressions are flattened (echoing similar concerns in LLM systems \citet{agarwal2025ai}). Speech AI tools can reduce and limit users' self-expression \cite{xu2025your} and the uniqueness of their avowed identities, reducing authorial voice and pushing users toward average or otherwise reductive speech representations. This is especially prominent in cases where people feel insecure about their skills or pressured to perform a particular identity (e.g. in AI-led job interviews), as they may surrender their cognitive load to the system~\cite{shaw2026thinking} and conform to a self-representation different from their avowed identity. More broadly, text suggestions in text-to-speech tools can reduce the expression of personal identity in speech output; this impact is especially salient for AAC users~\cite{valencia-2023-less, kane2017times, xu2025your}. For any user whose identity, culture, or communication style sits outside the dominant patterns, this averaging may be experienced as a form of erasure \cite{wenzel2024designing}.

Another issue for avowed identity regards how speech AI tools privilege Western cisgender voices. Current English speech synthesis systems are dominated by American and British English accents, and multilingual systems generally are dominated by high-resource languages \cite{xinyuan2025scalable, zhong2025accentbox, ogun20241000}. This means users from other linguistic backgrounds may get a voice that sounds nothing like them or even need to adapt a voice for another language (such as in \citet{callahan2023reconnecting}, where a Czech voice was used for the Native American Lakota), reinforcing linguistic privilege. Systems purporting to generate synthetic AI voices, if not carefully designed and evaluated, risk amplifying existing linguistic hierarchies and accent-based discrimination~\cite{michel2025bias}. This points to a representational ceiling that technical improvements to synthesis alone cannot solve: the barrier for the most underserved users lies not in generation quality but in the absence of voices that reflect their cultural register, linguistic identity, and sense of self in the first place.

Even with a diverse and representative speech dataset, creating an accurate measure of identity alignment remains difficult due to the nuances of voice identity utilization. For example, AAC users may consider how long they've been using, and grown to identify with, a given voice, how many (possibly more famous) others use the same voice, and the situational use of different voices \cite{preece2024making} alongside the more typical language, accent, age, gender, and expressiveness. Voice identity in AAC is not fixed at the moment of selection; it accumulates over time through use, accruing social history, relational meaning, and familiarity that can make a long-used voice feel like one's own even when it was never a good fit --- and make a better-fitting voice feel wrong precisely because it is unfamiliar~\cite{weinberg2026voice}. At the same time, AAC users do not inhabit a single stable identity; self-presentation is often context-dependent, with users navigating different registers across relationships, settings, and life roles~\cite{weinberg2026voice, weinberg2026robot}.
These fluid identity needs surface in concrete voice choices that go beyond the typical dimensions of language, accent, age, and gender. AAC is part of gender expression, so it is part of transgender expression \cite{Tuttleturtle2020gender,zisk2021male,danielescu2023creating,weinberg2026voice}; likewise, a disabled person may want an audibly disabled voice, and may prefer a synthetic one that most listeners find easier to understand than their embodied speech~\cite{preece2024making,weinberg2026voice}. Creating a metric that accurately encompasses the nuances of voice identity alignment remains an unaddressed challenge in speech AI.

When AAC users cannot access voices that match their preferences on all axes, they are forced to pick from a selection of incorrect voices to determine a (possibly situational) least wrong option \cite{preece2024making,Tuttleturtle2020gender,zisk2021male,weinberg2026voice}. This may lead to unexpected choices in prioritization, such as attempting to sidestep synthetic voices \cite{weinberg2026robot,sellwood2024imagining}, a multilingual AAC user selecting a voice on the basis of prosody matching most closely with their identity, even if sacrificing accurate pronunciation \cite{callahan2023reconnecting}, or using voices of different genders between different languages \cite{zisk2025trying}. Linguistic minorities may need to adapt voices from other languages and dialects, and some non-binary AAC users have needed to adapt voices from other genders \cite{Tuttleturtle2020gender, zisk2021male} as non-binary AAC voices have often not been available \cite{danielescu2023creating}. Across all of these cases, the choice of voice is never purely aesthetic --- it is a negotiation between what the system can offer, what the user's social context demands, and what identity the user is trying to project or protect.  

When users cannot aptly choose their avowed identity, they impact their ascribed identity and how others may perceive them by their synthetic voice. Listeners readily assign race, gender, and regional identity to synthetic and voice-assistant speech, and those assignments carry the same social consequences --- credibility judgments, warmth, and competence --- that accompany perceived identity in human speech \cite{holliday2021perception, holliday2023siri}. Importantly,  how listeners perceive and socially categorize a given synthetic voice is independent of whether the speaker behind it recognizes it as theirs. This means a voice can be a poor match on one axis and a reasonable match on another; a voice a user identifies with may still be miscategorized by listeners, and a voice built to be broadly intelligible or `neutral' to listeners may fail to feel like the user's own. The gap between these perspectives is measurable in adjacent settings. Models trained on third-party emotion labels transfer poorly to self-reported emotion \cite{el2026you}, and closing that gap requires models personalized to the individual, since systems fit to consensus annotations cannot be assumed to recover what a speaker reports about themselves \cite{tavernor2026personal}. A user may thus select a voice that feels like their own and still be misread by everyone who hears it, which means voice selection is a negotiation across two audiences whose criteria need not agree.

Listener perceptions have a significant impact on how the quality of synthetic voices is evaluated, which means a voice can score well on technical tests but still feel wrong to the actual user. Current speech synthesis evaluation relies on listeners scoring or assessing strangers' voices, for which they have no perceptual baseline. Recent work has shown that listeners can distinguish between synthetic speech of themselves and their own recordings with high accuracy, while highlighting accent and prosody as two main factors where speech synthesis models are perceptually different \cite{sanchez2026cui}. This suggests that, when speech synthesis is evaluated with speakers unknown to the listeners, we miss the opportunity of a potentially more nuanced understanding of creating a TTS model that works for users.
\begin{solutionsketch}[2B]{\tSystem\tMetric\tUser}
Designing better speech AI tools requires giving users accessible ways to articulate, explore, and refine identity-linked vocal preferences over time~\cite{ahmed2022app, netzorg2024speech, weinberg2026voice}. This includes creating an accurate metric for voice identity alignment, and adjusting such metrics to account for both avowed and ascribed voice identity.
\end{solutionsketch}

\section*{Problem 3: Speech AI metrics measure the wrong constructs.}
The sub-problems below share a shape: something easy to count is treated as the thing we actually care about. We count how many words a system got wrong, and call that a measure of whether the transcript served the speaker (Problem 3A). We measure how close a synthesized voice sounds to a recording, and call that a measure of whether the voice fits its user (Problem 3B). We average scores across unrelated benchmarks, and call that a measure of how good a system is (Problem 3C). We test on short isolated clips, and call that a measure of understanding (Problem 3D). Each number may accurately quantify its chosen proxy, but none fully captures the construct for which it is commonly taken to stand.

Goodhart's law is an old adage, commonly stated as `\textit{When a measure becomes a target, it ceases to be a good measure}.' Once a benchmark is widely adopted, optimization pressure concentrates on the narrow slice of behavior it happens to score, and the metric drifts from the construct it was meant to measure to the proxy itself~\cite{goodhart1984problems,manheim2018categorizing,el2024goodhart}. We hold this is not a marginal concern in speech AI, but the dominant failure mode of how the field currently evaluates its own progress.

\subsection*{Problem 3A: Speech recognition metrics obscure real-world harm.}
Speech recognition systems have traditionally been evaluated using quantitative metrics such as word error rate (WER), which quantifies normalized edit distance between model predictions and reference transcriptions, or match error rate (MER), which measures the errors between aligned words in the reference and model output~\cite{morris-wer-2004}. While these metrics provide a standardized way to compare model performance, they are known to be inadequate in capturing the downstream user impact of transcription errors. For example, WER treats all substitutions equally, failing to distinguish between harmful substitutions (e.g., profanity, identity-related slurs, or threatening language) and semantically benign errors such as homophones~\cite{wu2019design}. To address some of these limitations, more  quantitative metrics have been introduced to quantify the semantic distance between the prediction and the reference, including BERTScore~\cite{zhang2020bertscore} and SemScore~\cite{aynetdinov2024semscore}. 

Going beyond redesigning error metrics and reference transcripts, evaluation must also directly measure the consequences of errors in situated interaction. For example, can users reliably tell the difference between a speech-to-text systems with average WERs of 5\% versus 10\%? How do similar recognition errors affect users differently across contexts such as medical documentation, workplace communication, or casual texting? How do different groups interpret and react to the transcription failures differently? These questions suggest that the limitation of current benchmarking metrics are not solely technical, but also contextual and social. Existing metrics have allowed speech AI developers to abstract away the lived experience of interacting with speech AI systems, obscuring how system failures may reinforce marginalization, exclusion, or psychological harm.

Previous HCI research has empirically explored some of these questions. For example, \citet{wenzel-2023-microaggressors} showed that voice assistant errors produce disparate psychological impacts for Black and white users. Relatedly, ~\citet{kadoma2026lost} found that erroneous transcription of speech reduces both speaker and content evaluations, potentially disproportionately impacting minoritized speakers. Other work has documented how fluent transcription of stuttered speech can reinforce fluency dominance and erase the speaker's identity as a person who stutters~\cite{li2024reenvisioning}, while in contrast, aphasia patients reported preferring their transcriptions to be written in the `cleanest' form, erasing stutters~\cite{mei2026addressing}. We argue that speech AI evaluation must move beyond aggregate recognition accuracy and toward user-centered evaluation frameworks that explicitly measure contextualized harms and lived user experience. 

Developing such evaluation frameworks requires community-centered research that identifies which forms of recognition failure matter most to affected users. Studies involving people with aphasia and people who stutter have shown that generative ASR systems hallucinate more frequently for these populations, sometimes generating degrading, threatening, or violent content that causes tangible harms~\cite{Koenecke-2024-careless, sridhar2025jjj}. These findings call for new evaluation measures such as hallucination rate and harmful hallucination rate. Recent work with people who stutter further challenges normative assumptions embedded in conventional ASR evaluation~\cite{li2025collective, sridhar2025jjj}. Existing transcription practices often prioritize `intended speech' over verbatim speech production, implicitly framing disfluencies as errors to be removed rather than meaningful aspects of communication~\cite{teleki24_interspeech, radford2022whisper, choi26_speechspectrum, range-wer}, impacting SpeechLLMs \cite{teleki26_dres}. The stuttering community expressed a desire for greater control over whether and how speech disfluencies are preserved in transcripts, motivating proposals such as verbatim WER, which evaluates recognition performance against verbatim rather than normalized transcripts~\cite{li2025collective, sridhar2025jjj}. 

Developing community-centered evaluation metrics also requires new datasets and annotation methodologies. Existing ASR training and benchmarking datasets routinely exclude so-called `non-verbal' audio segments and prioritize transcription of intended speech rather than faithful representation of speech production \cite{radford2022whisper, romana2024fluencybank}. As a result, speech blocks, repetitions, prolongations, and other forms of disfluency that frequently occur in certain groups, such as people who stutter, are often absent from existing datasets. These omissions not only limit model performance for marginalized speech communities, but also encode normative assumptions about what constitutes valid or desirable speech.

Finally, we also need to complement the benchmarking metrics, which are often calculated over clean, static benchmarking datasets, with direct measurement on user experience and sentiment in situated interactions. Prior work by \citet{wenzel-2023-microaggressors}, for instance, measured users' affective responses to voice assistants with varying error rates, including self assessments of self-consciousness and self-esteem. Such approaches directly examine the psychological and social impact of speech AI systems, closing the gap between static metrics and lived experience and user well-being. 

\begin{solutionsketch}[3A]{\tField\tSystem\tPartic}
Future work should focus on contextualizing performance benchmarks within users' lived experience. This necessitates conducting community-embedded research, and creating new datasets and annotation methods.
\end{solutionsketch}

\subsection*{Problem 3B: Personalized voice evaluations don't consider AAC user needs.}
The same distance between measured performance and lived experience opens on the synthesis side, where the available instruments are fewer and the AAC context is served by none of them.
There may be limited quantitative metrics for speech synthesis in general \cite{wagner2019speech}, and most speech synthesis is not designed for the AAC context in any case \cite{pullin2025discussing}. Whether designed for speech synthesis or adapted to speech synthesis from speech recognition, this limits the metrics used in AAC contexts, and we are not aware of any \textit{designed for} the AAC context. Metrics adapted from other areas of speech AI for use in the AAC space include character error rate, word error rate, the structural similarity index, and Speaker Encoder Cosine Similarity \cite{szekely2025voice,sanchez25_interspeech}. Character error rate, word error rate and structural similarity were used to evaluate the effectiveness of prompting an AAC system with dysarthric speech, possibly analogous to the automation of revoicing \cite{preece2024making,pullin2025discussing,sellwood2024imagining}.  Speaker Encoder Cosine Similarity was used to evaluate similarity to recordings for an AAC user who previously spoke and may also be appropriate to evaluate the similarity of synthesized voices before and after speech engine updates. However, there may not be an appropriate comparator when evaluating and selecting new voices for AAC users who never spoke orally \cite{preece2024making} or do not want an AAC voice that sounds like their embodied voice \cite{Tuttleturtle2020gender,zisk2021male}.

A broader family of objective speech evaluation metrics is also available, including intrusive quality measures such as PESQ 
(ITU-T P.862) and its successors \cite{rix2001pesq}, which score a synthesised utterance against a clean recording of the same content in the target voice. This is a requirement often inappropriate in an AAC context, since for a user who never spoke orally \cite{preece2024making} or who does not want their embodied voice \cite{Tuttleturtle2020gender, zisk2021male}, no such reference exists.
Reference-free predictors such as UTMOS \cite{saeki2022utmos} attempt to predict human preference. While they are non-intrusive, they inherit typical-speech and unfamiliar-listener assumptions instead.

The main limitation with current speech synthesis evaluation for AAC is that it follows standard speech synthesis evaluation methods, which rely on listeners' perceptions of voices that they are not familiar with, when familiarity has been shown to significantly change the perception of cloned voices \cite{Rosi25_chb}. Even though recent work has proposed a framework for evaluating personalized speech synthesis that aims for a more situationally framed, preference-based approach \cite{sanchez26_interspeech}, it is still a non-user centric evaluation. Although these evaluations may still provide insight into the technical performance of personalized speech synthesis, they do not capture what AAC users themselves want from their voices. As future work, an evaluation for personalised speech synthesis developed alongside AAC users is long overdue.  

\begin{solutionsketch}[3B]{\tBench\tUser}
We propose evaluating personalized voices with three types of listeners: (1) the user of the AAC system, (2) listeners familiar with the user of the voice, and (3) unfamiliar listeners to the user of the voice. For listeners (2) and (3), evaluations should be contextually-framed \cite{edlund24_interspeech} and mimic real-life use cases wherein the AAC user will communicate with listeners. Evaluating these voices from three distinct listeners can provide more insights, not only based on preference and fidelity to the AAC user, but also in terms of usability in different contexts (e.g., talking to your friend, or ordering a coffee).
\end{solutionsketch}

\subsection*{Problem 3C: Leaderboards prioritize single-number evaluation.}
The limitations in 3A and 3B share a structure: a single number stands in for a construct it partially captures, and the field then optimizes the number.

Widely used community leaderboards such as the HuggingFace Open ASR Leaderboard\footnote{\url{https://huggingface.co/spaces/hf-audio/open_asr_leaderboard}} reinforce this dynamic; a single average of Word Error Rates (WER\%) ranks systems across dozens of unrelated benchmarks and conditions. The resulting ranking becomes a reference point for model development, despite potentially reflecting no real deployment context in particular. It can also privilege organizations with the resources to optimize across many benchmarks simultaneously. Examples of other such proxy metrics and associated problems, apart from WER~\cite{phukon2025aligning} include Mean Opinion Scores (MOS)~\cite{lemaguer2024limits, omahony21_ssw}, speaker-similarity cosine distances~\cite{carbonneau2025analyzing}, and Multiple Choice Question Answering (MCQA) measures for bias~\cite{lin2024spoken}, as in NLP~\cite{parrish2022bbq,nadeem2021stereoset,jin-etal-2025-social}. 

These metrics underplay what users care about, \emph{e.g.}, intelligibility-in-context, voice naturalness and feeling heard, having one's voice and identity reflected back faithfully, whether AI responses are fair and useful to them --- or more domain-specific metrics, such as hospital preferences for AI medical scribes to transcribe patient notes using correct medical terminology and integrating seamlessly with electronic health record systems~\cite{koenecke2026perspective}.

We do not propose abandoning quantitative evaluation; we propose pluralizing it, adding complementing evaluations to it and being explicit about \textit{which metrics matter to whom, and for what decisions}, following \citet{blodgett2020language}.

\begin{solutionsketch}[3C]{\tBench\tFieldbuild\tPartic\tMetric}
We propose three commitments: (1) Treat every benchmark as a partial, situated metric and report robustness across voices, prompts, decoding settings, and task formats by default, not as an ablation, so that single-number leaderboards stop being a credible currency~\cite{bokkahallisatish2025dobias,bokkahalli2025voice,frisch-complicated-2026, ethayarajh2020utility, wang2024benchmark}.

(2) Pair every aggregate metric with at least one evaluation in which affected communities and end-users define and evaluate what forms of evidence matter, how success should be evaluated and how the resulting outcomes are interpreted. For instance, this could mean qualitative studies of a small set of deliberately diverse $N=1$\footnote{$N=1$ in the sense of highly-intersectional, complex users; consider, for example, a user with a unique set of health conditions, who can realistically say \textit{``there is no one like me.''}} cases. Recognize and design evaluations for realistic scenarios, human variability and design measurements keeping in mind the people who currently fail or are failed by the system~\cite{bokkahalli2026voice, bondi2021envisioning, zee2024group, justin2026eyaa, mei2026addressing}, and accept that we are intervening in a complex adaptive system rather than tuning a static objective.

(3) Build evaluation \textit{interfaces}, not just scores, that let stakeholders interrogate model behavior directly under controlled perturbations of voice~\cite{bokkahalli2025hear, li2025mind, wei2026bias}. Such interfaces resist value capture because they refuse to compress the evaluation down to a single optimizable number.
\end{solutionsketch}

\subsection*{Problem 3D: Speech AI models have a limited understanding of context -- in both the Conversation Analysis (CA) sense and the model window sense.}
A major technical limitation of current speech AI is its limited understanding of conversational context. Under the hood, systems often separate literal words from vocal cues like tone and emotion~\cite{huang2026tone}, a disconnect made worse by processing audio in short, isolated clips rather than holding a longer memory of the interaction~\cite{chaichana2026extending}. Consequently, models fail to recognize how broader social settings alter meaning; the exact same text and vocal tension convey vastly different intents in a formal office meeting versus casual friendly banter~\cite{vinciarelli2009social}. This contextual blindness spills over into tasks like machine translation, where correctly translating can require awareness of the speaker's personality, social context and environment~\cite{voita2019good}. Ultimately, because these models are evaluated using rigid tests that ignore context, high accuracy scores on curated datasets hide a severe measurement gap: their real-world inability to grasp the deeply context-dependent nature of human interaction~\cite{zhang2025wildspeech}. Transitioning towards scenario-specific evaluation metrics offers a promising path to resolving these blind spots. By developing benchmarks that measure performance within defined social and environmental contexts, future models can successfully bridge this gap, toward achieving contextual intelligence.
 
\begin{solutionsketch}[3D]{\tBench\tSystem}
To address these limitations, we suggest prioritizing more realistic benchmarks that measure how well a model responds to intent and social environment, moving beyond isolated transcription scores. This shift in evaluation can, in turn, drive architectural improvements like unified acoustic-semantic embeddings, long-term conversational memory buffers, and the explicit injection of situational metadata.
\end{solutionsketch}

\section*{A Call for Structural Change}
The three problems follow from a division of labor: one community holds the instruments that make speech systems measurable, the other the vocabulary that makes their social consequences describable, and neither suffices alone for a modality that encodes identity in every utterance. Whose voice is heard, on whose terms, and with what consequences for identity are questions that broader corpora and disaggregated metrics can inform but cannot settle by themselves --- which is why nearly every sketch we offer carries tags from more than one methodological family, and why measuring what these systems do to people requires benchmarks and field studies, metrics and co-design, in combination.

AAC anchors the argument because it is where the stakes are least deniable: a user whose voice is the system's output, whose identity is negotiated through a menu of imperfect options, and who is judged by listeners on the result. What AAC makes vivid is present wherever speech AI is deployed. Much remains unsettled --- what diversity should mean here, whether instruments imported from text and vision can be repaired or need replacing, who holds authority over a voice once it has entered a model --- and each of these requires both communities in the room.

Getting them into the room is a matter of incentives, and those incentives are set by reviewers and program committees. We therefore address them directly. The routes available to the kind of work we describe are uneven: *ACL venues admit position work into the main track and route speech submissions through a dedicated speech area, while ICASSP and Interspeech call for technical papers in a single format with no contribution type for a critical argument or reflection, so a claim about what the field should measure reaches those reviewers once it has been instantiated as results. FAccT, CHI, and ASSETS accept critique and construction alike, at some remove from the practitioners who maintain the benchmarks in question. Organizers hold the levers that close this distance: seating area chairs with cross-community expertise, recruiting reviewer pools that can assess both kinds of contribution, creating evaluation tracks and special themes where this work has a natural home, and stating in calls for papers that methodological pluralism is expected. Where those levers go unpulled, the sketches we offer will be written by people whose venues will not reward them for it, and the technical and sociotechnical communities will continue to publish past each other. 

We close with an invitation to build the shared vocabulary this work depends on. At the scale of a community, that means joint workshops and boundary-spanning papers of every kind: position papers such as this one, and equally the mixed-methods studies written to be read on both sides of the divide. At the scale of a single researcher it is smaller, and available without anyone's permission: write the next paper with the other half of the room in mind. Cite across the divide. Recruit a co-author trained differently. Change accumulates one paper at a time.

\section*{Acknowledgments}
AI tools were used to assist with the language of the paper.
We thank everyone who attended and participated in the SpeechAI4All Workshop at CHI `26 \cite{10.1145/3772363.3778768}.

NC is part-funded by the National Institute for Health and Care Research (NIHR) Biomedical Research Centre (BRC): Maudsley. The views expressed are those of the author(s) and not necessarily those of the NIHR or the Department of Health and Social Care.

SHBS was partially supported by the Wallenberg AI, Autonomous Systems and Software Program (WASP) funded by the Knut and Alice Wallenberg Foundation.

\section*{Limitations}
This is a position paper, and our contributions are solution sketches: early ideas meant to open cross-community work, not finished methods or evaluated systems. We are excited about the directions they point toward, and we offer them in the expectation that the shape each one finally takes will be worked out by the researchers who take them up — refined, recombined, and put in contact with real users, real data, and real constraints. We see the sketches as starting points, and we look forward to what emerges as different communities build on them.

Our critique draws on AAC as its anchoring case. Given AAC users' reliance on speech AI, we believe this case study may be an instantiation of the curb-cut effect, wherein optimizing for users who rely on these technologies most can yield benefits for the majority of users \cite{blackwell2017curb}. Nonetheless, AAC users are not a monolith, and lessons drawn from this setting will not transfer cleanly to every underserved speaker. Future research may explore the needs of the other communities we invoke more deeply, such as people who stutter, non-binary and transgender voice users, multilingual speakers, speakers of low-resource languages, older adults, children, and others. Our account of `both communities' likewise flattens real internal heterogeneity within the technical and socio-technical venues we name, and the boundary between them is more porous than our framing sometimes suggests; we use this lens as we believe cross-community collaboration is central to progress. This framing helps to highlight each community's strength and weakness, and opportunities to assist one another.

\section*{Ethical Considerations}
Several of our sketches carry risks that must be addressed as that work unfolds. Problem 1, in particular, admits physiological and facial sensing as opt-in inputs to an expressive model; such sensing is invasive by construction and raises acute concerns around consent, surveillance, and data governance, especially for AAC users whose communication technology is already deeply entangled with their identity and daily autonomy. The personalization directions in Problem 2 raise parallel questions about who owns a cloned voice, who controls the model trained on someone's speech, and how a system built to sound like a person might come to constrain or misrepresent them. Our treatment of Problem 3 also assumes an actor with the resources to evaluate however it chooses. In practice, much of the existing evaluation apparatus is already difficult for smaller speech AI companies to commit resources to partaking in, and our critique of what those instruments measure says nothing about who is in a position to run them. A fuller account would treat the metrics we discuss as constrained by cost as well as by construct. We flag these tensions throughout, but we do not resolve them, and we caution against reading any sketch as a mandate to collect intimate data absent community-defined safeguards.


\begin{CJK*}{UTF8}{gbsn} 
\bibliography{custom}

@article{holliday2025gender,
  title={Gender and racial bias issues in a commercial “tone of voice” analysis system},
  author={Holliday, Nicole R and Reed, Paul E},
  journal={PloS one},
  volume={20},
  number={2},
  pages={e0314470},
  year={2025},
  publisher={Public Library of Science San Francisco, CA USA}
}

@inproceedings{murad2022voice,
  title={“Voice-First Interfaces in a GUI-First Design World”: Barriers and Opportunities to Supporting VUI Designers On-the-Job},
  author={Murad, Christine and Tasnim, Humaira and Munteanu, Cosmin},
  booktitle={Proceedings of the 4th Conference on Conversational User Interfaces},
  pages={1--10},
  year={2022}
}

@inproceedings{michel2025bias,
  title={“It’s not a representation of me”: Examining Accent Bias and Digital Exclusion in Synthetic AI Voice Services},
  author={Michel, Shira and Kaur, Sufi and Gillespie, Sarah Elizabeth and Gleason, Jeffrey and Wilson, Christo and Ghosh, Avijit},
  booktitle={Proceedings of the 2025 ACM Conference on Fairness, Accountability, and Transparency},
  pages={228--245},
  year={2025}
}

@article{weinberg2026voice,
  title={Me, Myself, and My Voice: Exploring Cultural and Linguistic Identity in AAC AI-generated Voices},
  author={Weinberg, Tobias M and Lewis, Aaleyah and Penuela, Ricardo E Gonzalez and Hong, Weicong and Mankoff, Jennifer and Roumen, Thijs},
  journal={arXiv preprint arXiv:2605.24337},
  year={2026}
}

@inproceedings{kane2017times,
  title={" At times avuncular and cantankerous, with the reflexes of a mongoose" Understanding Self-Expression through Augmentative and Alternative Communication Devices},
  author={Kane, Shaun K and Morris, Meredith Ringel and Paradiso, Ann and Campbell, Jon},
  booktitle={Proceedings of the 2017 acm conference on computer supported cooperative work and social computing},
  pages={1166--1179},
  year={2017}
}

@inproceedings{weinberg2025why,
  title={Why So Serious? Exploring Timely Humorous Comments in AAC Through AI-Powered Interfaces},
  author={Weinberg, Tobias M and Kadoma, Kowe and Gonzalez Penuela, Ricardo E and Valencia, Stephanie and Roumen, Thijs},
  booktitle={Proceedings of the 2025 CHI Conference on Human Factors in Computing Systems},
  pages={1--19},
  year={2025}
}

@inproceedings{jurafsky1997automatic,
  title={Automatic detection of discourse structure for speech recognition and understanding},
  author={Jurafsky, Daniel and Bates, Rebecca and Coccaro, Noah and Martin, Rachel and Meteer, Marie and Ries, Klaus and Shriberg, Elizabeth and Stolcke, Andreas and Taylor, Paul and Van Ess-Dykema, Carol},
  booktitle={1997 IEEE Workshop on Automatic Speech Recognition and Understanding Proceedings},
  pages={88--95},
  year={1997},
  organization={IEEE}
}

@inproceedings{sin2023cui,
  title={CUI@ CHI: Inclusive Design of CUIs Across Modalities and Mobilities},
  author={Sin, Jaisie and Candello, Heloisa and Clark, Leigh and Cowan, Benjamin R and Lee, Minha and Munteanu, Cosmin and Porcheron, Martin and V{\"o}lkel, Sarah Theres and Branham, Stacy and Brewer, Robin N and others},
  booktitle={Extended Abstracts of the 2023 CHI Conference on Human Factors in Computing Systems},
  pages={1--5},
  year={2023}
}

@inproceedings{leoni-etal-2024-solving,
    title = "Solving Failure Modes in the Creation of Trustworthy Language Technologies",
    author = "Leoni, Gianna  and
      Steven, Lee  and
      Keith, T{\={u}}reiti  and
      Mahelona, Keoni  and
      Jones, Peter-Lucas  and
      Duncan, Suzanne",
    editor = "Melero, Maite  and
      Sakti, Sakriani  and
      Soria, Claudia",
    booktitle = "Proceedings of the 3rd Annual Meeting of the Special Interest Group on Under-resourced Languages @ LREC-COLING 2024",
    month = may,
    year = "2024",
    address = "Torino, Italia",
    publisher = "ELRA and ICCL",
    url = "https://aclanthology.org/2024.sigul-1.39/",
    pages = "325--330"
}

@inproceedings{sin2021digital,
  title={Digital design marginalization: New perspectives on designing inclusive interfaces},
  author={Sin, Jaisie and L. Franz, Rachel and Munteanu, Cosmin and Barbosa Neves, Barbara},
  booktitle={Proceedings of the 2021 CHI Conference on Human Factors in Computing Systems},
  pages={1--11},
  year={2021}
}

@inproceedings{sin2025beyond,
  title={Beyond Deficit-Based Design: Re-imagining Co-Creation Approaches with Equity-Denied Groups},
  author={Sin, Jaisie and Conte, Sho and Munteanu, Cosmin and Maitland, Jules and Nurain, Novia and Sarcar, Sayan and Zhao, Wei},
  booktitle={Proceedings of the Extended Abstracts of the CHI Conference on Human Factors in Computing Systems},
  pages={1--5},
  year={2025}
}

@article{schneider2022multilingualism,
  title={Multilingualism and AI: The regimentation of language in the age of digital capitalism},
  author={Schneider, Britta},
  journal={Signs and Society},
  volume={10},
  number={3},
  pages={362--387},
  year={2022},
  publisher={Cambridge University Press \& Assessment}
}

@inproceedings{wagner2019speech,
  title={Speech synthesis evaluation—state-of-the-art assessment and suggestion for a novel research program},
  author={Wagner, Petra and Beskow, Jonas and Betz, Simon and Edlund, Jens and Gustafson, Joakim and Eje Henter, Gustav and Le Maguer, S{\'e}bastien and Malisz, Zofia and Sz{\'e}kely, {\'E}va and T{\aa}nnander, Christina and others},
  booktitle={Proceedings of the 10th Speech Synthesis Workshop (SSW10)},
  year={2019}
}

@article{pullin2025discussing,
  title={Discussing future {AAC}: technology, interactions and ownership},
  author={Pullin, Graham and Williams, Kevin and Sellwood, Darryl and Preece, Jamie and Sullivan, Emma and Allan, Meredith and Hutchinson, Todd and Brown, Katie and Tams-Gray, Fin and Higginbotham, Jeff},
  journal={Augmentative and Alternative Communication},
  pages={1--7},
  year={2025},
  publisher={Taylor \& Francis}
}

@inproceedings{szekely2025voice,
  title={Voice Reconstruction through Large-Scale TTS Models: Comparing Zero-Shot and Fine-tuning Approaches to Personalise TTS in Assistive Communication},
  author={Sz{\'e}kely, {\'E}va and Mihajlik, P{\'e}ter and K{\'a}d{\'a}r, M{\'a}t{\'e} Soma and T{\'o}th, L{\'a}szl{\'o}},
  booktitle={Proc. Interspeech},
  pages={2735--2739},
  year={2025}
}

@incollection{zisk2025trying,
    author = {Zisk, Alyssa Hillary},
    editor= {Kapila, Ruchi and Chan, Mai Ling},
    title = {Trying to say 怎么fucking了: code-mixing with {AAC} (augmentative and alternative communication)},
    booktitle = {Community Engagement for Speech-Language Pathology: From Lived Experience to Clinical Expertise},
    publisher = {Cognella Publishing},
    year = {Forthcoming},
}

@article{sellwood2024imagining,
  title={Imagining alternative futures with augmentative and alternative communication: a manifesto},
  author={Sellwood, Darryl and McLeod, Lateef and Williams, Kevin and Brown, Katie and Pullin, Graham},
  journal={Medical Humanities},
  volume={50},
  number={4},
  pages={620--623},
  year={2024},
  publisher={Institute of Medical Ethics}
}

@article{hill2020deaf,
  title={Do deaf communities actually want sign language gloves?},
  author={Hill, Joseph},
  journal={Nature Electronics},
  volume={3},
  number={9},
  pages={512--513},
  year={2020},
  publisher={Nature Publishing Group UK London}
}

@article{zisk2021male,
    author = {Zisk, Alyssa Hillary},
    title = {“Why are you using a male voice?”},
    journal = {CommunicationFIRST Perspectives},
    year = 2021,
    url = {https://communicationfirst.org/why-are-you-using-a-male-voice/}
}

@conference{Tuttleturtle2020gender,
    author = {{Tuttleturtle (S. Fuller)}},
    booktitle = {{AAC} in the Cloud} ,
    title = {My {AAC} is Part of My Gender Presentation},
    year = 2020,
    url = {https://presenters.AACconference.com/videos/U1RVMVFUSXc}
}

@article{preece2024making,
  title={Making my voice and owning its future},
  author={Preece, Jamie and Sullivan, Emma and Tams-Gray, Fin and Pullin, Graham},
  journal={Medical Humanities},
  volume={50},
  number={4},
  pages={624--634},
  year={2024},
  publisher={Institute of Medical Ethics}
}

@article{callahan2023reconnecting,
  title={Reconnecting Indigenous language for a child using augmentative and alternative communication},
  author={Callahan, Janet and Hanson, Elizabeth K},
  journal={Language, Speech, and Hearing Services in Schools},
  volume={54},
  number={2},
  pages={387--394},
  year={2023},
  publisher={American Speech-Language-Hearing Association}
}

@inproceedings{weinberg2026robot,
  title={I, robot? exploring ultra-personalized ai-powered {AAC}; an autoethnographic account},
  author={Weinberg, Tobias M and Gonzalez Penuela, Ricardo E and Valencia, Stephanie and Roumen, Thijs},
  booktitle={Proceedings of the 2026 CHI Conference on Human Factors in Computing Systems},
  pages={1--16},
  year={2026},
  doi = {10.1145/3772318.3790310}
}

@inproceedings{teleki25_icwsm,
title = {Masculine Defaults via Gendered Discourse in Podcasts and Large Language Models},
author = {Maria Teleki and Xiangjue Dong and Haoran Liu and James Caverlee},
year = {2025},
booktitle = {ICWSM 2025}
}

@article{holliday2021perception,
  title={Perception in black and white: Effects of intonational variables and filtering conditions on sociolinguistic judgments with implications for ASR},
  author={Holliday, Nicole R},
  journal={Frontiers in Artificial Intelligence},
  volume={4},
  pages={642783},
  year={2021},
  publisher={Frontiers Media SA}
}

@article{holliday2023siri,
  title={Siri, you've changed! Acoustic properties and racialized judgments of voice assistants},
  author={Holliday, Nicole},
  journal={Frontiers in Communication},
  volume={8},
  pages={1116955},
  year={2023},
  publisher={Frontiers Media SA}
}

@inproceedings{choi2025fairness,
  title={Fairness of Automatic Speech Recognition: Looking Through a Philosophical Lens},
  author={Choi, Anna Seo Gyeong and Choi, Hoon},
  booktitle={Proceedings of the AAAI/ACM Conference on AI, Ethics, and Society},
  volume={8},
  number={1},
  pages={605--614},
  year={2025}
}

@book{costanza2020design,
  title={Design justice: Community-led practices to build the worlds we need},
  author={Costanza-Chock, Sasha},
  year={2020},
  publisher={MIT press}
}

@inproceedings{birhane2022power,
  title={Power to the people? Opportunities and challenges for participatory AI},
  author={Birhane, Abeba and Isaac, William and Prabhakaran, Vinodkumar and Diaz, Mark and Elish, Madeleine Clare and Gabriel, Iason and Mohamed, Shakir},
  booktitle={Proceedings of the 2nd ACM Conference on Equity and Access in Algorithms, Mechanisms, and Optimization},
  pages={1--8},
  year={2022}
}

@inproceedings{garcia2020no,
  title={No: Critical refusal as feminist data practice},
  author={Garcia, Patricia and Sutherland, Tonia and Cifor, Marika and Chan, Anita Say and Klein, Lauren and D'Ignazio, Catherine and Salehi, Niloufar},
  booktitle={Companion Publication of the 2020 Conference on Computer Supported Cooperative Work and Social Computing},
  pages={199--202},
  year={2020}
}

@inproceedings{teleki26_dres,
  title = {Conversational Speech Reveals Structural Robustness Failures in SpeechLLM Backbones},
  author = {Maria Teleki and Sai Janjur and Haoran Liu and Oliver Grabner and Ketan Verma and Thomas Docog and Xiangjue Dong and Lingfeng Shi and Cong Wang and Stephanie Birkelbach and Jason Kim and Yin Zhang and Éva Székely and James Caverlee},
  year = {2026},
}

@inproceedings{choi26_speechspectrum,
title = {SpeechSpectrum: A Linguistic Fidelity Spectrum for Accountable Speech-to-Text},
author = {Anna Seo Gyeong Choi and Maria Teleki and Miguel del Rio and James Caverlee and Corey Miller and Allison Koenecke},
year = {2026},
booktitle = {Preprint}
}

@inproceedings{range-wer,
title = {Beyond Single Ground Truth: Reference Monism as Epistemic Injustice in {ASR} Evaluation},
author = {Anna Seo Gyeong Choi and Maria Teleki and James Caverlee and Miguel del Rio and Corey Miller and Hoon Choi},
year = {2026},
booktitle = {Preprint}
}

@inproceedings{teleki24_interspeech,
title = {Comparing ASR Systems in the Context of Speech Disfluencies},
author = {Maria Teleki and Xiangjue Dong and Soohwan Kim and James Caverlee},
year = {2024},
booktitle = {Interspeech 2024},
pages = {4548--4552},
doi = {10.21437/Interspeech.2024-1270},
}

@article{goodhart1984problems,
  author  = {Goodhart, Charles A. E.},
  title   = {{Problems of Monetary Management: The U.K. Experience}},
  journal = {Papers in Monetary Economics},
  volume  = {1},
  pages   = {91--121},
  year    = {1984},
  publisher = {Reserve Bank of Australia}
}

@article{manheim2018categorizing,
  author  = {Manheim, David and Garrabrant, Scott},
  title   = {Categorizing Variants of Goodhart's Law},
  journal = {arXiv preprint arXiv:1803.04585},
  year    = {2018}
}

@article{el2024goodhart,
  title   = {On Goodhart's law, with an application to value alignment},
  author  = {El-Mhamdi, El-Mahdi and Hoang, L{\^e}-Nguy{\^e}n},
  journal = {arXiv preprint arXiv:2410.09638},
  year    = {2024}
}

@inproceedings{bokkahallisatish2025dobias,
  title={{Do Bias Benchmarks Generalise? Evidence from Voice-Based Evaluation of Gender Bias in SpeechLLMs}},
  author={Bokkahalli Satish, Shree Harsha and Henter, Gustav Eje and Sz{\'e}kely, {\'E}va},
  booktitle={Proc. ICASSP},
  pages={4566--4570},
  year={2026},
  organization={IEEE}
}

@inproceedings{bokkahalli2025hear,
  title={{Hear Me Out: Interactive evaluation and bias discovery platform for speech-to-speech conversational AI}},
  author={Bokkahalli Satish, Shree Harsha and Henter, Gustav Eje and Sz{\'e}kely, {\'E}va},
  booktitle={Proc. Interspeech},
  pages={2151--2152},
  year={2025},
  organization={International Speech Communication Association}
}

@article{bokkahalli2026voice,
  title={{The Voice Behind the Words: Quantifying Intersectional Bias in SpeechLLMs}},
  author={Bokkahalli Satish, Shree Harsha and Minixhofer, Christoph and Teleki, Maria and Caverlee, James and Klejch, Ond{\'L} and Bell, Peter and Henter, Gustav Eje and Sz{\'e}kely, {\'E}va},
  journal={arXiv preprint arXiv:2603.16941},
  year={2026}
}

@inproceedings{bokkahalli2025voice,
  title={{When Voice Matters: Evidence of Gender Disparity in Positional Bias of SpeechLLMs}},
  author={Bokkahalli Satish, Shree Harsha and Henter, Gustav Eje and Sz{\'e}kely, {\'E}va},
  booktitle={International Conference on Speech and Computer},
  pages={25--38},
  year={2025}
}

@book{levinson_interaction_2025,
    edition = {1},
    title = {The {Interaction} {Engine}: {Language} in {Social} {Life} and {Human} {Evolution}},
    copyright = {https://www.cambridge.org/core/terms},
    isbn = {978-1-009-57034-3 978-1-009-57032-9 978-1-009-57035-0},
    shorttitle = {The {Interaction} {Engine}},
    url = {https://www.cambridge.org/core/product/identifier/9781009570343/type/book},
    doi = {10.1017/9781009570343},
    urldate = {2026-04-27},
    publisher = {Cambridge University Press},
    author = {Levinson, Stephen C.},
    month = jun,
    year = {2025},
}

@article{goodwin_conversation_1990,
    title = {Conversation {Analysis}},
    volume = {19},
    url = {https://doi.org/10.1146/annurev.an.19.100190.001435},
    doi = {10.1146/annurev.an.19.100190.001435},
    number = {1},
    urldate = {2023-03-28},
    journal = {Annual Review of Anthropology},
    author = {Goodwin, Charles and Heritage, John},
    year = {1990},
    note = {\_eprint: https://doi.org/10.1146/annurev.an.19.100190.001435},
    pages = {283--307},
}

@article{goodwin_action_2000,
    title = {Action and embodiment within situated human interaction},
    volume = {32},
    issn = {0378-2166},
    url = {https://www.sciencedirect.com/science/article/pii/S037821669900096X},
    doi = {10.1016/S0378-2166(99)00096-X},
    language = {en},
    number = {10},
    urldate = {2021-02-16},
    journal = {Journal of Pragmatics},
    author = {Goodwin, Charles},
    month = sep,
    year = {2000},
    pages = {1489--1522},
}

@inproceedings{ibrahim_design_2018,
    address = {New York, NY, USA},
    series = {{CHI} '18},
    title = {Design {Opportunities} for {AAC} and {Children} with {Severe} {Speech} and {Physical} {Impairments}},
    copyright = {All rights reserved},
    isbn = {978-1-4503-5620-6},
    url = {https://dl.acm.org/citation.cfm?id=3173801},
    doi = {10.1145/3173574.3173801},
    urldate = {2018-05-17},
    booktitle = {Proceedings of the 2018 {CHI} {Conference} on {Human} {Factors} in {Computing} {Systems}},
    publisher = {ACM},
    author = {Ibrahim, Seray B. and Vasalou, Asimina and Clarke, Michael},
    year = {2018},
    pages = {227:1--227:13},
}

@inproceedings{valencia_conversational_2020,
    address = {New York, NY, USA},
    series = {{CHI} '20},
    title = {Conversational {Agency} in {Augmentative} and {Alternative} {Communication}},
    isbn = {978-1-4503-6708-0},
    url = {https://dl.acm.org/doi/10.1145/3313831.3376376},
    doi = {10.1145/3313831.3376376},
    urldate = {2023-03-28},
    booktitle = {Proceedings of the 2020 {CHI} {Conference} on {Human} {Factors} in {Computing} {Systems}},
    publisher = {Association for Computing Machinery},
    author = {Valencia, Stephanie and Pavel, Amy and Santa Maria, Jared and Yu, Seunga (Gloria) and Bigham, Jeffrey P. and Admoni, Henny},
    month = apr,
    year = {2020},
    pages = {1--12},
}

@inproceedings{weinberg_one_2025,
    address = {New York, NY, USA},
    series = {{ASSETS} '25},
    title = {One {Does} {Not} {Simply} ‘{Mm}-hmm’: {Exploring} {Backchanneling} in the {AAC} {Micro}-{Culture}},
    isbn = {979-8-4007-0676-9},
    shorttitle = {One {Does} {Not} {Simply} ‘{Mm}-hmm’},
    url = {https://dl.acm.org/doi/10.1145/3663547.3746381},
    doi = {10.1145/3663547.3746381},
    urldate = {2026-01-09},
    booktitle = {Proceedings of the 27th {International} {ACM} {SIGACCESS} {Conference} on {Computers} and {Accessibility}},
    publisher = {Association for Computing Machinery},
    author = {Weinberg, Tobias M and O'Connor, Claire and Gonzalez Penuela, Ricardo E. and Valencia, Stephanie and Roumen, Thijs},
    month = oct,
    year = {2025},
    pages = {1--14},
}

@article{higginbotham2002aac,
  title={{AAC} performance and usability issues: The effect of {AAC} technology on the communicative process},
  author={Higginbotham, D Jeffery and Caves, Kevin},
  journal={Assistive Technology},
  volume={14},
  number={1},
  pages={45--57},
  year={2002},
  publisher={Taylor \& Francis}
}

@book{clark_using_1996, 
    place={Cambridge}, 
    series={“Using” Linguistic Books}, 
    title={Using Language}, 
    publisher={Cambridge University Press}, 
    author={Clark, Herbert H.}, 
    year={1996}, 
    collection={“Using” Linguistic Books}
}

@article{bloch_understandability_2004,
    author = {Steven Bloch and Ray Wilkinson},
    title = {The Understandability of {AAC}: A Conversation Analysis Study of Acquired Dysarthria},
    journal = {Augmentative and Alternative Communication},
    volume = {20},
    number = {4},
    pages = {272--282},
    year = {2004},
    publisher = {Taylor \& Francis},
    doi = {10.1080/07434610400005614},
    URL = {https://doi.org/10.1080/07434610400005614}
}

@article{poplack_codeswitching_1980,
    author = {Poplack, Shana},
    year = {1980},
    month = {01},
    pages = {581-618},
    title = {Sometimes I’ll start a sentence in Spanish Y TERMINO EN ESPAÑOL: toward a typology of code-switching},
    volume = {18},
    journal = {Linguistics},
    doi = {10.1515/ling.1980.18.7-8.581}
}

@article{harrington_acoustic_2006,
  title={An acoustic analysis of ‘happy-tensing’in the Queen's Christmas broadcasts},
  author={Harrington, Jonathan},
  journal={Journal of Phonetics},
  volume={34},
  number={4},
  pages={439--457},
  year={2006},
  publisher={Elsevier}
}

@article{rosen_longitudinal_2012,
  title={Longitudinal change in dysarthria associated with Friedreich ataxia: a potential clinical endpoint},
  author={Rosen, Kristin M and Folker, Joanne E and Vogel, Adam P and Corben, Louise A and Murdoch, Bruce E and Delatycki, Martin B},
  journal={Journal of neurology},
  volume={259},
  number={11},
  pages={2471--2477},
  year={2012},
  publisher={Springer}
}

@article{bucholtz_identity_2005,
  title={Identity and interaction: A sociocultural linguistic approach},
  author={Bucholtz, Mary and Hall, Kira},
  journal={Discourse studies},
  volume={7},
  number={4-5},
  pages={585--614},
  year={2005},
  publisher={sage Publications London, Thousand Oaks, CA and New Delhi}
}

@article{koenecke2020racial,
  title={Racial disparities in automated speech recognition},
  author={Koenecke, Allison and Nam, Andrew and Lake, Emily and Nudell, Joe and Quartey, Minnie and Mengesha, Zion and Toups, Connor and Rickford, John R and Jurafsky, Dan and Goel, Sharad},
  journal={Proceedings of the national academy of sciences},
  volume={117},
  number={14},
  pages={7684--7689},
  year={2020},
  publisher={National Academy of Sciences}
}

@inproceedings{wu2025speech,
  title={Speech ai for all: Promoting accessibility, fairness, inclusivity, and equity},
  author={Wu, Shaomei and Wenzel, Kimi and Li, Jingjin and Li, Qisheng and Pradhan, Alisha and Kushalnagar, Raja and Lea, Colin and Koenecke, Allison and Vogler, Christian and Hasegawa-Johnson, Mark and others},
  booktitle={Proceedings of the Extended Abstracts of the CHI Conference on Human Factors in Computing Systems},
  pages={1--6},
  year={2025}
}

@inproceedings{blodgett2020language,
  title={Language (technology) is power: A critical survey of “bias” in {NLP}},
  author={Blodgett, Su Lin and Barocas, Solon and Daum{\'e} Iii, Hal and Wallach, Hanna},
  booktitle={Proceedings of the 58th annual meeting of the association for computational linguistics},
  pages={5454--5476},
  year={2020}
}

@article{eckert2019limits,
  title={The limits of meaning: Social indexicality, variation, and the cline of interiority},
  author={Eckert, Penelope},
  journal={Language},
  volume={95},
  number={4},
  pages={751--776},
  year={2019},
  publisher={Cambridge University Press}
}

@inproceedings{cunningham2025advancing,
  title={Advancing {NLP} Data Equity: Practitioner Responsibility and Accountability in {NLP} Data Practices},
  author={Cunningham, Jay L and Shao, Kevin Zhongyang and Pang, Rock Yuren and Mengist, Nathanael Elias},
  booktitle={Proceedings of the AAAI/ACM Conference on AI, Ethics, and Society},
  volume={8},
  pages={679--691},
  year={2025}
}

@article{foulkes2006social,
  title={The social life of phonetics and phonology},
  author={Foulkes, Paul and Docherty, Gerard},
  journal={Journal of phonetics},
  volume={34},
  number={4},
  pages={409--438},
  year={2006},
  publisher={Elsevier}
}

@article{eckert2008variation,
  title={Variation and the indexical field 1},
  author={Eckert, Penelope},
  journal={Journal of sociolinguistics},
  volume={12},
  number={4},
  pages={453--476},
  year={2008},
  publisher={Wiley Online Library}
}

@inproceedings{nguyen2025we,
  title={We Need to Measure Data Diversity in {NLP}—Better and Broader},
  author={Nguyen, Dong and Ploeger, Esther},
  booktitle={Proceedings of the 2025 Conference on Empirical Methods in Natural Language Processing},
  pages={8823--8832},
  year={2025}
}

@inproceedings{buolamwini2018gender,
  title={Gender shades: Intersectional accuracy disparities in commercial gender classification},
  author={Buolamwini, Joy and Gebru, Timnit},
  booktitle={Conference on fairness, accountability and transparency},
  pages={77--91},
  year={2018},
  organization={PMLR}
}

@inproceedings{szekely2025will,
  title={Will {AI} shape the way we speak? The emerging sociolinguistic influence of synthetic voices},
  author={Sz{\'e}kely, {\'E}va and Miniota, Jura and Hejn{\'a}, M{\'\i}{\v{s}}a Michaela},
  booktitle={Proceedings of the 15th International Workshop on Spoken Dialogue Systems Technology},
  pages={335--340},
  year={2025}
}

@article{fazelpour2022diversity,
  title={Diversity in sociotechnical machine learning systems},
  author={Fazelpour, Sina and De-Arteaga, Maria},
  journal={Big Data \& Society},
  volume={9},
  number={1},
  pages={20539517221082027},
  year={2022},
  publisher={SAGE Publications Sage UK: London, England}
}

@inproceedings{dheram2022toward,
  title={Toward Fairness in Speech Recognition: Discovery and mitigation of performance disparities},
  author={Dheram, Pranav and Ramakrishnan, Murugesan and Raju, Anirudh and Chen, I-Fan and King, Brian and Powell, Katherine and Saboowala, Melissa and Shetty, Karan and Stolcke, Andreas},
  booktitle={Proc. Interspeech},
  pages={1268--1272},
  year={2022}
}

@inproceedings{tobin2025towards,
  title={Towards a single {ASR} model that generalizes to disordered speech},
  author={Tobin, Jimmy and Tomanek, Katrin and Venugopalan, Subhashini},
  booktitle={Proc. ICASSP},
  pages={1--5},
  year={2025},
  organization={IEEE}
}

@inproceedings{tomanek2023analysis,
  title={An analysis of degenerating speech due to progressive dysarthria on {ASR} performance},
  author={Tomanek, Katrin and Seaver, Katie and Jiang, Pan-Pan and Cave, Richard and Harrell, Lauren and Green, Jordan R},
  booktitle={Proc. ICASSP},
  pages={1--5},
  year={2023},
  organization={IEEE}
}

@article{erete2018intersectional,
  title={An intersectional approach to designing in the margins},
  author={Erete, Sheena and Israni, Aarti and Dillahunt, Tawanna},
  journal={interactions},
  volume={25},
  number={3},
  pages={66--69},
  year={2018},
  publisher={ACM New York, NY, USA}
}

@inproceedings{liu2022model,
  title={Model-based approach for measuring the fairness in {ASR}},
  author={Liu, Zhe and Veliche, Irina-Elena and Peng, Fuchun},
  booktitle={Proc. ICASSP},
  pages={6532--6536},
  year={2022},
  organization={IEEE}
}

@inproceedings{wenzel-2023-microaggressors,
author = {Wenzel, Kimi and Devireddy, Nitya and Davison, Cam and Kaufman, Geoff},
title = {Can Voice Assistants Be Microaggressors? Cross-Race Psychological Responses to Failures of Automatic Speech Recognition},
year = {2023},
isbn = {9781450394215},
publisher = {Association for Computing Machinery},
address = {New York, NY, USA},
url = {https://doi.org/10.1145/3544548.3581357},
doi = {10.1145/3544548.3581357},
booktitle = {Proceedings of the 2023 CHI Conference on Human Factors in Computing Systems},
articleno = {109},
numpages = {14},
location = {Hamburg, Germany},
series = {CHI '23}
}

@inproceedings{li2024reenvisioning,
author = {Li, Jingjin and Wu, Shaomei and Leshed, Gilly},
title = {Re-envisioning Remote Meetings: Co-designing Inclusive and Empowering Videoconferencing with People Who Stutter},
year = {2024},
isbn = {9798400705830},
publisher = {Association for Computing Machinery},
address = {New York, NY, USA},
url = {https://doi.org/10.1145/3643834.3661533},
doi = {10.1145/3643834.3661533},
booktitle = {Proceedings of the 2024 ACM Designing Interactive Systems Conference},
pages = {1926–1941},
numpages = {16},
location = {Copenhagen, Denmark},
series = {DIS '24}
}

@inproceedings{li2025collective,
author = {Li, Jingjin and Li, Qisheng and Gong, Rong and Wang, Lezhi and Wu, Shaomei},
title = {Our Collective Voices: The Social and Technical Values of a Grassroots Chinese Stuttered Speech Dataset},
year = {2025},
isbn = {9798400714825},
publisher = {Association for Computing Machinery},
address = {New York, NY, USA},
url = {https://doi.org/10.1145/3715275.3732179},
doi = {10.1145/3715275.3732179},
booktitle = {Proceedings of the 2025 ACM Conference on Fairness, Accountability, and Transparency},
pages = {2768–2783},
numpages = {16},
location = {
},
series = {FAccT '25}
}

@inproceedings{Koenecke-2024-careless,
author = {Koenecke, Allison and Choi, Anna Seo Gyeong and Mei, Katelyn X. and Schellmann, Hilke and Sloane, Mona},
title = {Careless Whisper: Speech-to-Text Hallucination Harms},
year = {2024},
isbn = {9798400704505},
publisher = {Association for Computing Machinery},
address = {New York, NY, USA},
url = {https://doi.org/10.1145/3630106.3658996},
doi = {10.1145/3630106.3658996},
booktitle = {Proceedings of the 2024 ACM Conference on Fairness, Accountability, and Transparency},
pages = {1672–1681},
numpages = {10},
location = {Rio de Janeiro, Brazil},
series = {FAccT '24}
}

@misc{aynetdinov2024semscore,
      title={SemScore: Automated Evaluation of Instruction-Tuned LLMs based on Semantic Textual Similarity}, 
      author={Ansar Aynetdinov and Alan Akbik},
      year={2024},
      eprint={2401.17072},
      archivePrefix={arXiv},
      primaryClass={cs.CL},
      url={https://arxiv.org/abs/2401.17072}, 
}

@misc{zhang2020bertscore,
      title={BERTScore: Evaluating Text Generation with BERT}, 
      author={Tianyi Zhang and Varsha Kishore and Felix Wu and Kilian Q. Weinberger and Yoav Artzi},
      year={2020},
      eprint={1904.09675},
      archivePrefix={arXiv},
      primaryClass={cs.CL},
      url={https://arxiv.org/abs/1904.09675}, 
}

@inproceedings{sridhar2025jjj,
  title={J-j-j-just Stutter: Benchmarking Whisper’s Performance Disparities on Different Stuttering Patterns},
  author={Sridhar, Charan and Wu, Shaomei},
  booktitle={Proc. Interspeech},
  pages={3753--3757},
  year={2025}
}

@misc{radford2022whisper,
      title={Robust Speech Recognition via Large-Scale Weak Supervision}, 
      author={Alec Radford and Jong Wook Kim and Tao Xu and Greg Brockman and Christine McLeavey and Ilya Sutskever},
      year={2022},
      eprint={2212.04356},
      archivePrefix={arXiv},
      primaryClass={eess.AS},
      url={https://arxiv.org/abs/2212.04356}, 
}

@inproceedings{wu2019design,
  title={Design and evaluation of a social media writing support tool for people with dyslexia},
  author={Wu, Shaomei and Reynolds, Lindsay and Li, Xian and Guzm{\'a}n, Francisco},
  booktitle={Proceedings of the 2019 CHI Conference on Human Factors in Computing Systems},
  pages={1--14},
  year={2019}
}

@article{romana2024fluencybank,
  title={FluencyBank Timestamped: An updated data set for disfluency detection and automatic intended speech recognition},
  author={Romana, Amrit and Niu, Minxue and Perez, Matthew and Provost, Emily Mower},
  journal={Journal of Speech, Language, and Hearing Research},
  volume={67},
  number={11},
  pages={4203--4215},
  year={2024},
  publisher={American Speech-Language-Hearing Association}
}

@inproceedings{valencia-2023-less,
author = {Valencia, Stephanie and Cave, Richard and Kallarackal, Krystal and Seaver, Katie and Terry, Michael and Kane, Shaun K.},
title = {“The less I type, the better”: How AI Language Models can Enhance or Impede Communication for AAC Users},
year = {2023},
isbn = {9781450394215},
publisher = {Association for Computing Machinery},
address = {New York, NY, USA},
url = {https://doi.org/10.1145/3544548.3581560},
doi = {10.1145/3544548.3581560},
booktitle = {Proceedings of the 2023 CHI Conference on Human Factors in Computing Systems},
articleno = {830},
numpages = {14},
location = {Hamburg, Germany},
series = {CHI '23}
}

@inproceedings{wenzel2026speech,
  title={Speech AI for All: The What, How, and Who of Measurement},
  author={Wenzel, Kimi and Pradhan, Alisha and Teleki, Maria and Weinberg, Tobias M and Netzorg, Robin and Hillary Zisk, Alyssa and Choi, Anna Seo Gyeong and Li, Jingjin and Kushalnagar, Raja and Lea, Colin and others},
  booktitle={Proceedings of the Extended Abstracts of the 2026 CHI Conference on Human Factors in Computing Systems},
  pages={1--6},
  year={2026}
}

@inproceedings{danielescu2023creating,
  title={Creating inclusive voices for the 21st century: A non-binary text-to-speech for conversational assistants},
  author={Danielescu, Andreea and Horowit-Hendler, Sharone A and Pabst, Alexandria and Stewart, Kenneth Michael and Gallo, Eric M and Aylett, Matthew Peter},
  booktitle={Proceedings of the 2023 CHI Conference on Human Factors in Computing Systems},
  pages={1--17},
  year={2023}
}

@article{huang2026tone,
  title={When Tone and Words Disagree: Towards Robust Speech Emotion Recognition under Acoustic-Semantic Conflict},
  author={Huang, Dawei and Lv, Yongjie and Xiong, Ruijie and Jin, Chunxiang and Peng, Xiaojiang},
  journal={arXiv preprint arXiv:2601.04564},
  year={2026}
}

@inproceedings{chaichana2026extending,
  title={Extending Audio Context for Long-Form Understanding in Large Audio-Language Models},
  author={Chaichana, Yuatyong and Taveekitworachai, Pittawat and Sirichotedumrong, Warit and Manakul, Potsawee and Pipatanakul, Kunat},
  booktitle={Proceedings of the 19th Conference of the European Chapter of the Association for Computational Linguistics (Volume 1: Long Papers)},
  pages={6046--6066},
  year={2026}
}

@article{vinciarelli2009social,
  title={Social signal processing: Survey of an emerging domain},
  author={Vinciarelli, Alessandro and Pantic, Maja and Bourlard, Herv{\'e}},
  journal={Image and vision computing},
  volume={27},
  number={12},
  pages={1743--1759},
  year={2009},
  publisher={Elsevier}
}

@inproceedings{voita2019good,
  title={When a good translation is wrong in context: Context-aware machine translation improves on deixis, ellipsis, and lexical cohesion},
  author={Voita, Elena and Sennrich, Rico and Titov, Ivan},
  booktitle={Proceedings of the 57th annual meeting of the association for computational linguistics},
  pages={1198--1212},
  year={2019}
}

@article{zhang2025wildspeech,
  title={WildSpeech-Bench: Benchmarking End-to-End SpeechLLMs in the Wild},
  author={Zhang, Linhao and Zhang, Jian and Lei, Bokai and Wu, Chuhan and Liu, Aiwei and Jia, Wei and Zhou, Xiao},
  journal={arXiv preprint arXiv:2506.21875},
  year={2025}

}

@article{shaw2026thinking,
  title={Thinking-Fast, Slow, and Artificial: How AI is Reshaping Human Reasoning and the Rise of Cognitive Surrender},
  author={Shaw, Steven D and Nave, Gideon},
  journal={Available at SSRN 6097646},
  year={2026}
}

@misc{linkedinposts2025juttatreviranus,
  author    = {Treviranus, Jutta},
  title     = {We need new research metrics if we are to address disparity.},
  year      = {2025},
  howpublished = {LinkedIn Post},
  url = {https://www.linkedin.com/posts/juttatreviranus_nserc-equity-diversity-and-inclusion-share-7404296252747759617-Jis0/},
  note      = {Accessed: 2026-05-02},
}

@article{treviranus2018three,
  title={The three dimensions of inclusive design: Part one},
  author={Treviranus, Jutta},
  journal={Medium},
  year={2018}}

@book{franklin2014ursula,
  title={Ursula Franklin speaks: Thoughts and afterthoughts},
  author={Franklin, Ursula Martius},
  year={2014},
  publisher={McGill-Queen's Press-MQUP}
}

@inproceedings{sanchez2026cui,
  title={When Text-to-Speech Speaks in Your Voice: A Study on Public Perception},
  author={Sanchez, Ariadna and Zhong, Jinzuomu and Deligianni, Artemis and Ross, Alice and King, Simon},
  booktitle={To appear at Proceedings of the 7th ACM Conference on Conversational User Interfaces},
  year={2026}
}

@inproceedings{sanchez25_interspeech,
  title     = {{Can We Reconstruct a Dysarthric Voice with the Large Speech Model Parler TTS?}},
  author    = {Ariadna Sanchez and Simon King},
  year      = {2025},
  booktitle = {{Proc. Interspeech}},
  pages     = {4138--4142},
  doi       = {10.21437/Interspeech.2025-2679},
  issn      = {2958-1796},
}

@inproceedings{sanchez26_interspeech,
  title     = {{An Evaluation Framework for Text-to-Speech Voice Reconstruction}},
  author    = {Ariadna Sanchez and Christoph Minixhofer and Korin Richmond and Ondrej Klejch and Peter Bell and Simon King},
  year      = {2026},
  booktitle = {{To appear at Interspeech 2026}},
}

@inproceedings{edlund24_interspeech,
  title     = {{Assessing the impact of contextual framing on subjective TTS quality}},
  author    = {Jens Edlund and Christina Tånnander and Sébastien {Le Maguer} and Petra Wagner},
  year      = {2024},
  booktitle = {{Proc. Interspeech}},
  pages     = {1205--1209},
  doi       = {10.21437/Interspeech.2024-781},
  issn      = {2958-1796},
}

@article{Rosi25_chb,
title = {Perception and social evaluation of cloned and recorded voices: Effects of familiarity and self-relevance},
journal = {Computers in Human Behavior: Artificial Humans},
volume = {4},
year = {2025},
issn = {2949-8821},
doi = {10.1016/j.chbah.2025.100143},
author = {Victor Rosi and Emma Soopramanien and Carolyn McGettigan},
}

@article{saeki2022utmos,
  title={{UTMOS}: Utokyo-sarulab system for voicemos challenge 2022},
  author={Saeki, Takaaki and Xin, Detai and Nakata, Wataru and Koriyama, Tomoki and Takamichi, Shinnosuke and Saruwatari, Hiroshi},
  journal={arXiv preprint arXiv:2204.02152},
  year={2022}
}

@inproceedings{rix2001pesq,
  title={Perceptual evaluation of speech quality (PESQ)-a new method for speech quality assessment of telephone networks and codecs},
  author={Rix, Antony W and Beerends, John G and Hollier, Michael P and Hekstra, Andries P},
  booktitle={IEEE international conference on acoustics, speech, and signal processing},
  volume={2},
  pages={749--752},
  year={2001},
}

@misc{MDC2026linkedin,
  author       = {Mozilla Data Collective},
  title        = {{Get a sneak preview of Mozilla Data Collective’s community compensation feature!}},
  howpublished = {LinkedIn Post},
  year         = {2026},
  month        = jul,
  day          = {9},
  note         = {Accessed: 2026-07-09},
  url          = {https://www.linkedin.com/posts/datastewardship-responsibleai-datagovernance-share-7480953723725258752-MPu1/?utm_source=share&utm_medium=member_desktop&rcm=ACoAABPV8zcBXmoSZA9H1RIrbNpiTZ3iaQLlIE0}
}

@inproceedings{morris-wer-2004,
  title={From WER and RIL to MER and WIL: improved evaluation measures for connected speech recognition.},
  author={Morris, Andrew Cameron and Maier, Viktoria and Green, Phil D},
  booktitle={Interspeech},
  number={4-8},
  pages={2004},
  year={2004},
  doi={10.21437/Interspeech.2004-668}
}

@inproceedings{frisch-complicated-2026,
    author = {Blade Frisch and Will Wade and Dylan Gaines and Michelle Kinsella and Betts Peters and Tamara Broderick and Keith Vertanen},
    title = {It's Complicated: On the Design and Evaluation of AI-Powered AAC Interfaces},
    booktitle = {Speech AI for All: The What, How, and Who of Measurement Workshop at the CHI Conference on Human Factors in Computing Systems},
    year = {2026},
    doi = {10.48550/arXiv.2606.24854}
}

@inproceedings{parrish2022bbq,
  title={BBQ: A hand-built bias benchmark for question answering},
  author={Parrish, Alicia and Chen, Angelica and Nangia, Nikita and Padmakumar, Vishakh and Phang, Jason and Thompson, Jana and Htut, Phu Mon and Bowman, Samuel R},
  booktitle={Findings of the Association for Computational Linguistics: ACL 2022},
  pages={2086--2105},
  year={2022}
}

@inproceedings{nadeem2021stereoset,
  title={StereoSet: Measuring stereotypical bias in pretrained language models},
  author={Nadeem, Moin and Bethke, Anna and Reddy, Siva},
  booktitle={Proceedings of the 59th annual meeting of the association for computational linguistics and the 11th international joint conference on natural language processing (volume 1: long papers)},
  pages={5356--5371},
  year={2021}
}

@inproceedings{lin2024spoken,
  title={Spoken stereoset: on evaluating social bias toward speaker in speech large language models},
  author={Lin, Yi-Cheng and Chen, Wei-Chih and Lee, Hung-yi},
  booktitle={2024 IEEE Spoken Language Technology Workshop (SLT)},
  pages={871--878},
  year={2024},
  organization={IEEE}
}

@inproceedings{ethayarajh2020utility,
  title={Utility is in the eye of the user: A critique of NLP leaderboards},
  author={Ethayarajh, Kawin and Jurafsky, Dan},
  booktitle={Proceedings of the 2020 Conference on Empirical Methods in Natural Language Processing (EMNLP)},
  pages={4846--4853},
  year={2020}
}

@inproceedings{bondi2021envisioning,
  title={Envisioning communities: a participatory approach towards AI for social good},
  author={Bondi, Elizabeth and Xu, Lily and Acosta-Navas, Diana and Killian, Jackson A},
  booktitle={Proceedings of the 2021 AAAI/ACM Conference on AI, Ethics, and Society},
  pages={425--436},
  year={2021}
}

@inproceedings{li2025mind,
  title={Mind the Gap: Static and Interactive Evaluations of Large Audio Models},
  author={Li, Minzhi and Held, William Barr and Ryan, Michael J and Pipatanakul, Kunat and Manakul, Potsawee and Zhu, Hao and Yang, Diyi},
  booktitle={Proceedings of the 63rd Annual Meeting of the Association for Computational Linguistics (Volume 1: Long Papers)},
  pages={8749--8766},
  year={2025}
}

@inproceedings{zee2024group,
  title={Group fairness in multilingual speech recognition models},
  author={Zee, Anna and Zee, Marc and S{\o}gaard, Anders},
  booktitle={Findings of the Association for Computational Linguistics: NAACL 2024},
  pages={2213--2226},
  year={2024}
}

@article{wei2026bias,
  title={Bias in the Ear of the Listener: Assessing Sensitivity in Audio Language Models Across Linguistic, Demographic, and Positional Variations},
  author={Wei, Sheng-Lun and Liao, Yu-Ling and Chang, Yen-Hua and Huang, Hen-Hsen and Chen, Hsin-Hsi},
  journal={arXiv preprint arXiv:2602.01030},
  year={2026}
}

@inproceedings{justin2026eyaa,
  title={Eyaa-Tom 26, Yodi-Mantissa and Lom Bench: A Community Benchmark for TTS in Local Languages},
  author={Justin, Bakoubolo Essowe and Essuman, Catherine Nana Nyaah and Agbobli, Messan and Kansiwer, Ahoefa and Doumeyan, Eli Jean and Pato, Julie and Timibe, Notou Your and Agossou, Emile KOGBEDJI and Bakouya, Guedela},
  booktitle={Proceedings of the 7th Workshop on African Natural Language Processing (AfricaNLP 2026)},
  pages={264--270},
  year={2026}
}

@inproceedings{mei2026addressing,
  title={Addressing Auditing Pitfalls in Automatic Speech Recognition Technologies: A Case Study of People with Aphasia},
  author={Mei, Katelyn X and Choi, Anna Seo Gyeong and Schellmann, Hilke and Sloane, Mona and Koenecke, Allison},
  booktitle={The 2026 ACM Conference on Fairness, Accountability, and Transparency},
  pages={3422--3465},
  year={2026}
}

@inproceedings{papakyriakopoulos2023augmented,
  title={Augmented datasheets for speech datasets and ethical decision-making},
  author={Papakyriakopoulos, Orestis and Choi, Anna Seo Gyeong and Thong, William and Zhao, Dora and Andrews, Jerone and Bourke, Rebecca and Xiang, Alice and Koenecke, Allison},
  booktitle={Proceedings of the 2023 ACM Conference on Fairness, Accountability, and Transparency},
  pages={881--904},
  year={2023}
}

@inproceedings{kadoma2026lost,
  title={Lost in Transcription: Subtitle Errors in Automatic Speech Recognition Reduce Speaker and Content Evaluations},
  author={Kadoma, Kowe and Shrivastava, Priyal and Naaman, Mor},
  booktitle={Proceedings of the 2026 CHI Conference on Human Factors in Computing Systems},
  pages={1--11},
  year={2026}
}

@article{wang2024benchmark,
  title={Benchmark suites instead of leaderboards for evaluating AI fairness},
  author={Wang, Angelina and Hertzmann, Aaron and Russakovsky, Olga},
  journal={Patterns},
  volume={5},
  number={11},
  year={2024},
  publisher={Elsevier}
}

@article{ng2026end,
  title={An end-to-end overview of clinical speech ai},
  author={Ng, Si-Ioi and Xu, Lingfeng and Siegert, Ingo and Cummins, Nicholas and Benway, Nina R and Liss, Julie and Berisha, Visar},
  journal={IEEE Transactions on Audio, Speech and Language Processing},
  year={2026},
  publisher={IEEE}
}

@article{arora2025landscape,
  title={On the landscape of spoken language models: A comprehensive survey},
  author={Arora, Siddhant and Chang, Kai-Wei and Chien, Chung-Ming and Peng, Yifan and Wu, Haibin and Adi, Yossi and Dupoux, Emmanuel and Lee, Hung-Yi and Livescu, Karen and Watanabe, Shinji},
  journal={arXiv preprint arXiv:2504.08528},
  year={2025}
}

@inproceedings{cui2025recent,
  title={Recent advances in speech language models: A survey},
  author={Cui, Wenqian and Yu, Dianzhi and Jiao, Xiaoqi and Meng, Ziqiao and Zhang, Guangyan and Wang, Qichao and Guo, Steven Y and King, Irwin},
  booktitle={Proceedings of the 63rd Annual Meeting of the Association for Computational Linguistics (Volume 1: Long Papers)},
  pages={13943--13970},
  year={2025}
}

@article{yang2025large,
  title={When large language models meet speech: A survey on integration approaches},
  author={Yang, Zhengdong and Shimizu, Shuichiro and Yu, Yahan and Chu, Chenhui},
  journal={Findings of the Association for Computational Linguistics: ACL 2025},
  pages={20298--20315},
  year={2025}
}

@inproceedings{tavernor2026personal,
  title={It Is Personal: The Importance of Personalization for Recognizing Self-Reported Emotion},
  author={Tavernor, James and Provost, Emily Mower},
  booktitle={ICASSP 2026-2026 IEEE International Conference on Acoustics, Speech and Signal Processing (ICASSP)},
  pages={16202--16206},
  year={2026},
  organization={IEEE}
}

@article{el2026you,
  title={What You Feel Is Not What They See: On Predicting Self-Reported Emotion from Third-Party Observer Labels},
  author={El-Tawil, Yara and Sampath, Aneesha and Provost, Emily Mower},
  journal={arXiv preprint arXiv:2601.21130},
  year={2026}
}

@article{carbonneau2025analyzing,
  title={Analyzing and Improving Speaker Similarity Assessment for Speech Synthesis},
  author={Carbonneau, Marc-Andr{\'e} and van Niekerk, Benjamin and Seut{\'e}, Hugo and Letendre, Jean-Philippe and Kamper, Herman and Za{\"\i}di, Julian},
  journal={arXiv preprint arXiv:2507.02176},
  year={2025}
}

@inproceedings{jin-etal-2025-social,
  title     = {Social Bias Benchmark for Generation: A Comparison of
               Generation and QA-Based Evaluations},
  author    = {Jin, Jiho and Kang, Woosung and Myung, Junho and
               Oh, Alice},
  booktitle = {Findings of the Association for Computational
               Linguistics: ACL 2025},
  pages     = {11215--11228},
  year      = {2025},
  doi       = {10.18653/v1/2025.findings-acl.585}
}

@article{lemaguer2024limits,
  title   = {The Limits of the Mean Opinion Score for Speech
             Synthesis Evaluation},
  author  = {Le Maguer, S{\'e}bastien and King, Simon and Harte, Naomi},
  journal = {Computer Speech \& Language},
  volume  = {84},
  pages   = {101577},
  year    = {2024},
  doi     = {10.1016/j.csl.2023.101577}
}

@inproceedings{omahony21_ssw,
  title     = {Factors Affecting the Evaluation of Synthetic Speech
               in Context},
  author    = {O'Mahony, Johannah and Oplustil-Gallegos, Pilar and
               Lai, Catherine and King, Simon},
  booktitle = {11th ISCA Speech Synthesis Workshop},
  pages     = {148--153},
  year      = {2021},
  doi       = {10.21437/SSW.2021-26}
}

@article{phukon2025aligning,
  title={Aligning asr evaluation with human and llm judgments: Intelligibility metrics using phonetic, semantic, and nli approaches},
  author={Phukon, Bornali and Zheng, Xiuwen and Hasegawa-Johnson, Mark},
  journal={arXiv preprint arXiv:2506.16528},
  year={2025}
}

@inproceedings{ibrahim_before_2026,
    address = {New York, NY, USA},
    series = {{CHI} {EA} '26},
    title = {Before the {Technological} {Fix}: {Scoping} {AI} and {AAC} for {Social} {Futures}},
    isbn = {979-8-4007-2281-3},
    shorttitle = {Before the {Technological} {Fix}},
    url = {https://dl.acm.org/doi/10.1145/3772363.3798436},
    doi = {10.1145/3772363.3798436},
    urldate = {2026-07-20},
    booktitle = {Proceedings of the {Extended} {Abstracts} of the 2026 {CHI} {Conference} on {Human} {Factors} in {Computing} {Systems}},
    publisher = {Association for Computing Machinery},
    author = {Ibrahim, Seray B and Griffiths, Tom and Clarke, Michael and Judge, Simon and Slovak, Petr and Pullin, Graham and Higginbotham, Jeff},
    month = apr,
    year = {2026},
    pages = {1--10},
}

@inproceedings{bennett_interdependence_2018,
    address = {New York, NY, USA},
    series = {{ASSETS} '18},
    title = {Interdependence as a {Frame} for {Assistive} {Technology} {Research} and {Design}},
    isbn = {978-1-4503-5650-3},
    url = {https://doi.org/10.1145/3234695.3236348},
    doi = {10.1145/3234695.3236348},
    urldate = {2021-03-05},
    booktitle = {Proceedings of the 20th {International} {ACM} {SIGACCESS} {Conference} on {Computers} and {Accessibility}},
    publisher = {Association for Computing Machinery},
    author = {Bennett, Cynthia L. and Brady, Erin and Branham, Stacy M.},
    month = oct,
    year = {2018},
    pages = {161--173},
}

@inproceedings{delgado_participatory_2023,
    address = {New York, NY, USA},
    series = {{EAAMO} '23},
    title = {The {Participatory} {Turn} in {AI} {Design}: {Theoretical} {Foundations} and the {Current} {State} of {Practice}},
    isbn = {979-8-4007-0381-2},
    shorttitle = {The {Participatory} {Turn} in {AI} {Design}},
    url = {https://dl.acm.org/doi/10.1145/3617694.3623261},
    doi = {10.1145/3617694.3623261},
    urldate = {2026-07-20},
    booktitle = {Proceedings of the 3rd {ACM} {Conference} on {Equity} and {Access} in {Algorithms}, {Mechanisms}, and {Optimization}},
    publisher = {Association for Computing Machinery},
    author = {Delgado, Fernando and Yang, Stephen and Madaio, Michael and Yang, Qian},
    month = oct,
    year = {2023},
    pages = {1--23},
}

@inproceedings{agarwal2025ai,
  title={Ai suggestions homogenize writing toward western styles and diminish cultural nuances},
  author={Agarwal, Dhruv and Naaman, Mor and Vashistha, Aditya},
  booktitle={Proceedings of the 2025 CHI conference on human factors in computing systems},
  pages={1--21},
  year={2025}
}

@inproceedings{koenecke2026perspective,
  title = 	 {Perspective: Listening to Users when Auditing Medical AI Scribes},
  author =       {Koenecke, Allison and Nunez, John-Jose and Rameau, Ana{\"i}s and Chen, Irene Y.},
  booktitle = 	 {Proceedings of the Fifth Machine Learning for Health Symposium},
  pages = 	 {1619--1631},
  year = 	 {2026},
  editor = 	 {Argaw, Peniel and Zhang, Haoran and Jabbour, Sarah and Chandak, Payal and Ji, Jerry and Mukherjee, Sumit and Salaudeen, Olawale and Chang, Trenton and Healey, Elizabeth and Gröger, Fabian and Adibi, Amin and Hegselmann, Stefan and Wild, Benjamin and Noori, Ayush},
  volume = 	 {297},
  series = 	 {Proceedings of Machine Learning Research},
  month = 	 {13--14 Dec},
  publisher =    {PMLR},
  eprint = 	 {https://proceedings.mlr.press/v297/koenecke26a.html}
}

@article{xu2025your,
  title={Your voice is your voice: Supporting Self-expression through Speech Generation and LLMs in Augmented and Alternative Communication},
  author={Xu, Yiwen and Chakraborti, Monideep and Zhang, Tianyi and Eng, Katelyn and Mohan, Aanchan and Prpa, Mirjana},
  journal={arXiv preprint arXiv:2503.17479},
  year={2025}
}

@book{martin2010intercultural,
  title={Intercultural communication in contexts},
  author={Martin, Judith N and Nakayama, Thomas K},
  year={2010},
  publisher={United States: The McGraw-Hill Companies}
}

@article{rincon2021speaking,
  title={Speaking from experience: Trans/non-binary requirements for voice-activated AI},
  author={Rinc{\'o}n, Cami and Keyes, Os and Cath, Corinne},
  journal={Proceedings of the ACM on human-computer interaction},
  volume={5},
  number={CSCW1},
  pages={1--27},
  year={2021},
  publisher={ACM New York, NY, USA}
}

@inproceedings{nekoto2020participatory,
  title={Participatory research for low-resourced machine translation: A case study in african languages},
  author={Nekoto, Wilhelmina and Marivate, Vukosi and Matsila, Tshinondiwa and Fasubaa, Timi and Fagbohungbe, Taiwo and Akinola, Solomon Oluwole and Muhammad, Shamsuddeen Hassan and Kabenamualu, Salomon Kabongo and Osei, Salomey and Sackey, Freshia and others},
  booktitle={Findings of the Association for Computational Linguistics: EMNLP 2020},
  pages={2144--2160},
  year={2020}
}

@article{netzorg2024speech,
  title={Speech after gender: a trans-feminine perspective on next steps for speech science and technology},
  author={Netzorg, Robin and Cote, Alyssa and Koshin, Sumi and Garoute, Klo Vivienne and Anumanchipalli, Gopala Krishna},
  journal={arXiv preprint arXiv:2407.07235},
  year={2024}
}

@article{rojas2020does,
  title={How does our voice change as we age? A systematic review and meta-analysis of acoustic and perceptual voice data from healthy adults over 50 years of age},
  author={Rojas, Sandra and Kefalianos, Elaina and Vogel, Adam},
  journal={Journal of Speech, Language, and Hearing Research},
  volume={63},
  number={2},
  pages={533--551},
  year={2020},
  publisher={American Speech-Language-Hearing Association}
}

@inproceedings{blodgett2021stereotyping,
  title={Stereotyping Norwegian salmon: An inventory of pitfalls in fairness benchmark datasets},
  author={Blodgett, Su Lin and Lopez, Gilsinia and Olteanu, Alexandra and Sim, Robert and Wallach, Hanna},
  booktitle={Proceedings of the 59th Annual Meeting of the Association for Computational Linguistics and the 11th International Joint Conference on Natural Language Processing (Volume 1: Long Papers)},
  pages={1004--1015},
  year={2021}
}

@article{ahmed2022app,
  title={‘This app can help you change your voice’: Authenticity and authority in mobile applications for transgender voice training},
  author={Ahmed, Alex A and Kim, Levin and Hoffmann, Anna L},
  journal={Convergence},
  volume={28},
  number={5},
  pages={1283--1302},
  year={2022},
  publisher={SAGE Publications Sage UK: London, England}
}

@inproceedings{goldfarb2021intrinsic,
  title={Intrinsic bias metrics do not correlate with application bias},
  author={Goldfarb-Tarrant, Seraphina and Marchant, Rebecca and S{\'a}nchez, Ricardo Mu{\~n}oz and Pandya, Mugdha and Lopez, Adam},
  booktitle={Proceedings of the 59th Annual Meeting of the Association for Computational Linguistics and the 11th International Joint Conference on Natural Language Processing (Volume 1: Long Papers)},
  pages={1926--1940},
  year={2021}
}

@inproceedings{bowman2021will,
  title={What will it take to fix benchmarking in natural language understanding?},
  author={Bowman, Samuel and Dahl, George},
  booktitle={Proceedings of the 2021 Conference of the North American Chapter of the Association for Computational Linguistics: Human Language Technologies},
  pages={4843--4855},
  year={2021}
}

@article{xinyuan2025scalable,
  title={Scalable Controllable Accented TTS},
  author={Xinyuan, Henry Li and Cai, Zexin and Garg, Ashi and Duh, Kevin and Garc{\'\i}a-Perera, Leibny Paola and Khudanpur, Sanjeev and Andrews, Nicholas and Wiesner, Matthew},
  journal={arXiv preprint arXiv:2508.07426},
  year={2025}
}

@article{ogun20241000,
  title={1000 african voices: Advancing inclusive multi-speaker multi-accent speech synthesis},
  author={Ogun, Sewade and Owodunni, Abraham T and Olatunji, Tobi and Alese, Eniola and Oladimeji, Babatunde and Afonja, Tejumade and Olaleye, Kayode and Etori, Naome A and Adewumi, Tosin},
  journal={arXiv preprint arXiv:2406.11727},
  year={2024}
}

@inproceedings{zhong2025accentbox,
  title={AccentBox: Towards High-Fidelity Zero-Shot Accent Generation},
  author={Zhong, Jinzuomu and Richmond, Korin and Su, Zhiba and Sun, Siqi},
  booktitle={ICASSP 2025-2025 IEEE International Conference on Acoustics, Speech and Signal Processing (ICASSP)},
  pages={1--5},
  year={2025},
  organization={IEEE}
}

@article{blackwell2017curb,
  title={The curb-cut effect},
  author={Blackwell, Angela Glover},
  journal={Stanford Social Innovation Review},
  volume={15},
  number={1},
  pages={28--33},
  year={2017},
  publisher={Stanford: Stanford Center on Philanthropy and Civil Society}
}

@inproceedings{li2025govern,
  title={Govern With, Not For: Understanding the Stuttering Community’s Preferences and Goals for Speech AI Data Governance in the US and China},
  author={Li, Jingjin and Liu, Peiyao and Lietz, Rebecca and Tang, Ningjing and Su, Norman Makoto and Wu, Shaomei},
  booktitle={Proceedings of the AAAI/ACM Conference on AI, Ethics, and Society},
  volume={8},
  number={2},
  pages={1548--1560},
  year={2025}
}

@inproceedings{wenzel2024designing,
  title={Designing for harm reduction: Communication repair for multicultural users' voice interactions},
  author={Wenzel, Kimi and Kaufman, Geoff},
  booktitle={Proceedings of the 2024 CHI Conference on Human Factors in Computing Systems},
  pages={1--17},
  year={2024}
}

@inproceedings{10.1145/3772363.3778768,
author = {Wenzel, Kimi and Pradhan, Alisha and Teleki, Maria and Weinberg, Tobias M and Netzorg, Robin and Hillary Zisk, Alyssa and Choi, Anna Seo Gyeong and Li, Jingjin and Kushalnagar, Raja and Lea, Colin and Glasser, Abraham and Vogler, Christian and Brown, Ly Xundefinednzh\`{e}n M. Zhundefinedngs\={u}n and Ratner, Nan Bernstein and Koenecke, Allison and Nakamura, Karen and Wu, Shaomei},
title = {Speech AI for All: The What, How, and Who of Measurement},
year = {2026},
isbn = {9798400722813},
publisher = {Association for Computing Machinery},
address = {New York, NY, USA},
url = {https://doi.org/10.1145/3772363.3778768},
doi = {10.1145/3772363.3778768},
booktitle = {Proceedings of the Extended Abstracts of the 2026 CHI Conference on Human Factors in Computing Systems},
articleno = {969},
numpages = {6},
location = {
},
series = {CHI EA '26}
}
\end{CJK*}

\appendix

\section{Contributor Tags}
\label{appendix-contributor-tags}
Each solution sketch in the main text carries one or more \emph{contributor tags} that signal the kind of work the sketch invites, so that readers can scan for the sketches where their own methods apply. The seven tags fall into four color families: \emph{empirical study} (\iUser, \iField), \emph{measurement} (\iMetric, \iBench), \emph{system building} (\iSystem), and \emph{community practice} (\iPartic, \iFieldbuild). Most sketches span more than one: a single piece of work can, for instance, build a system and evaluate it with a user study. Here, we define each tag:

\newcommand{\tagentry}[2]{\vspace{5pt}\noindent#1\par\vspace{2pt}\noindent\ignorespaces #2\par}

\tagentry{\iUser\ \textbf{User Studies}}{Controlled, participant-facing empirical work: perception and listening tests, lab experiments, surveys, and other structured evaluations of how people perceive or respond to a system.}

\tagentry{\iField\ \textbf{Field Studies}}{Situated, in-context empirical work in real settings: interviews, ethnography, lived-experience accounts, and longitudinal or community-embedded observation.}

\tagentry{\iMetric\ \textbf{Metric Creation}}{Designing and validating a new measure or instrument: formalizing a construct and showing it captures what it claims to.} 

\tagentry{\iBench\ \textbf{Benchmarking}}{Building, running, and reporting evaluations against a measure: datasets, protocols, robustness reporting, and leaderboard practice.}

\tagentry{\iSystem\ \textbf{System Building}}{Designing and implementing models, architectures, interfaces, datasets, or pipelines.}

\tagentry{\iPartic\ \textbf{Participatory Design}}{Work done \emph{with} affected communities: co-design, data governance, ownership, and community-defined success criteria.}

\tagentry{\iFieldbuild\ \textbf{Field-Building}}{Work that changes how the field itself evaluates and rewards progress: cross-community infrastructure such as shared evaluation tracks, joint workshops, boundary-spanning position papers, and shifts in what the field treats as credible evidence.}

\section{Inclusive Design Practices}
If we are serious about reducing inequity and supporting meaningful innovation, we need to rethink what we measure. Alternative metrics should focus on the extent to which efforts reach and make a substantive difference for people facing the greatest barriers to inclusion, and the degree to which projects engage with complex, high-effort challenges that may take longer but have the potential for deeper, systemic impact. This reframing aligns with critiques by Jutta Treviranus  \cite{linkedinposts2025juttatreviranus}, who argues that prevailing evaluation systems undermine inclusive innovation by rewarding ease and scalability over necessity and equity.

Several overlapping traditions make a related argument from different starting points: design justice centers the leadership and needs of communities most harmed by design processes \cite{costanza2020design}, data feminism names how supposedly neutral data practices encode existing power hierarchies \cite{garcia2020no}, and participatory AI calls for affected communities to shape system design and evaluation directly rather than being consulted after the fact \cite{birhane2022power}. 
We draw here specifically on Treviranus's \cite{treviranus2018three} inclusive design framework, which advances an approach grounded in recognizing and designing for human variability, rather than conformity to an assumed norm. It emphasizes open, transparent, and participatory processes, including co-design with individuals who are often excluded or underserved by conventional design practices and technologies as defined by Ursula Franklin, ``\textit{Technology involves organizations, procedures, symbols, new words, equations, and most of all, Mindsets}.''
\cite{franklin2014ursula}. Situated within complex adaptive systems, Inclusive Design and wider situated design frameworks \cite[e.g.,][]{bennett_interdependence_2018, costanza2020design, delgado_participatory_2023, birhane2022power} understand design outcomes as emergent, contextualized, and interdependent, challenging the assumption that they can be optimized through standardization or efficiency-driven models. These frameworks call for alternative evaluation criteria: examining the extent to which technologies meaningfully engage those facing the greatest barriers to inclusion, and the degree to which projects address complex, systemic challenges with the potential for transformative impact. In this way, inclusive design not only reorients design practice but also redefines what counts as success and what mindset we apply to technology. 

\section{Exclusionary Practices}
Often, deficit-based design sees disability as something to \textit{fix} --- with technology's help \cite{sin2025beyond}. The problem is how pervasive uninclusive designs, i.e., designs that perpetuate exclusion of certain communities, introduce marginalization in which excluded communities experience deeper marginalization due to the increased lack of access; this can be, for instance, introduced inability to access healthcare apps or social media, which can be detrimental when considering enlarging social and psychological exclusion over time, a process termed as digital design marginalization \cite{sin2021digital}. While the problem is noted and inclusive practices are understood to be important \cite{sin2023cui,wu2025speech,wenzel2026speech}, there are additional hurdles, such as voice-first design principles being difficult to identify and implement, as VUI designers often rely on their prior GUI design experience for voice-based interaction design \cite{murad2022voice}.

\end{document}